\documentclass[a4paper,11pt]{article}
\usepackage{jheppub}
\usepackage{graphicx}
\usepackage{amsmath,amssymb,amsfonts}
\usepackage{xspace}
\usepackage{subfig}
\usepackage{float}
\usepackage{color}
\usepackage{xcolor}
\usepackage{bbding}
\usepackage[utf8]{inputenc}
\usepackage{commath}
\usepackage{slashed}
\usepackage{fancyvrb}
\usepackage[utf8]{inputenc}
\usepackage[english]{babel}
\usepackage{orcidlink}
\usepackage{cancel}
\usepackage{hyperref}
\usepackage{tabularray}
\UseTblrLibrary{booktabs}
\usepackage{gensymb}
\usepackage{scalerel}
\usepackage[titletoc]{appendix}

 \title{Gravitational Wave Probe of Singlet-Doublet Dark Matter Induced Radiative Neutrino Mass}

\author[a]{Ujjal Kumar Dey,$^{\orcidlink{https://orcid.org/0000-0002-9620-7561}}$,}
\emailAdd{ujjal@iiserbpr.ac.in}

\author[a]{Santu Kumar Manna,}
\emailAdd{santuk23@iiserbpr.ac.in}

\author[b]{Partha Kumar Paul$^{\orcidlink{https://orcid.org/0000-0002-9107-5635}}$,}
\emailAdd{ph22resch11012@iith.ac.in}

\author[b]{Sujit Kumar Sahoo$^{\orcidlink{https://orcid.org/0000-0002-9014-933X}}$,}
\emailAdd{ph21resch11008@iith.ac.in}

\author[b]{and Narendra Sahu$^{\orcidlink{https://orcid.org/0000-0002-9675-0484}}$}
\emailAdd{nsahu@phy.iith.ac.in}

\affiliation[a]{Department of Physical Sciences, Indian Institute of Science Education and Research
Berhampur, Ganjam, Odisha, 760003, India}
\affiliation[b]{Department of Physics, Indian Institute of Technology Hyderabad, Kandi, Telangana-502285, India.}

\abstract{We investigate an one loop radiative neutrino mass model, where the loop particles, notably a singlet fermion ($\chi$), a doublet fermion ($\Psi$) and three generations of singlet scalars ($\phi_i, i=\{1,2,3\}$) are assumed to be odd under an additional $\mathcal{Z}_2$-symmetry. In this setup, the singlet fermion mixes with the neutral component of the doublet to give rise singlet-doublet Majorana dark matter. The addition of $\mathcal{Z}_2$ odd scalars in the model provides rich phenomenological implications. We find that the quartic interaction terms between the SM Higgs and $\phi_i$s play a significant role in modifying the scalar potential to have a first-order phase transition (FOPT) leading to observable gravitational waves (GWs) spectra. We also examine the non-trivial role played by the singlet-doublet fermion DM and the scalars in loop-induced neutrino mass, $(g-2)_\mu$, and lepton flavor violation. We find that the model is predictive due to the combined constraints and can be verified at different terrestrial experiments. }

\keywords{Cosmology of Theories BSM, Models for Dark Matter, Particle Nature of Dark Matter, Phase Transitions in the Early Universe}

\arxivnumber{2511.19386}

\begin{document}
	
\maketitle
\flushbottom

%%%%%%%%%%%%%%%%%%%%%%%%%%%%%%%%%%%%%%%%%%%%%%%%%%%%%%%%%%%%%%%%%%%%%%%%%%%%%%%%%%%%%%%%%%
\section{Introduction}\label{sec:intro}

The Standard Model (SM) of particle physics is extremely successful in describing the masses of elementary particles and the interactions among them. However, it fails to explain the origin of the tiny sub-eV mass of neutrinos, which is necessary to explain the phenomenon of neutrino oscillation \cite{Super-Kamiokande:1998kpq,SNO:2001kpb,DoubleChooz:2011ymz,DayaBay:2012fng,RENO:2012mkc}. The simplest way to explain the neutrino mass is via the so-called \textit{seesaw} mechanism, which mainly includes the type-I \cite{Minkowski:1977sc, Gell-Mann:1979vob, Mohapatra:1979ia, Schechter:1980gr}, -II \cite{Mohapatra:1980yp, Lazarides:1980nt, Wetterich:1981bx, Schechter:1981cv, Ma:1998dx}, and -III \cite{Foot:1988aq} \textit{seesaw} at tree level. These mechanisms work at a very high scale, leaving their experimental verifiability beyond the scope of the present colliders. Instead, one can consider explaining the neutrino mass at the loop level \cite{Zee:1980ai,Ma:2006km,Mohapatra:2022tgb,Fraser:2014yha}. The advantage of considering such a model is that the mass scales of the particles running in the loop can be realized at the TeV scale, such that one can study their signatures at various experiments.

In this work, we consider a one-loop realization of the neutrino mass by extending the SM with a vector-like fermion doublet $\Psi$, one singlet fermion $\chi$, and three generations of singlet scalars $\phi_i (i=\{1,2,3\})$ \cite{Fraser:2014yha, Konar:2020wvl, Borah:2022zim}. All of these particles are odd under an imposed $\mathcal{Z}_2$ symmetry and constitute a dark sector. The lightest of these particles serves as the dark matter. The mass scale of all these particles can vary from the GeV to the TeV scale. This extension naturally generates sub-eV-scale masses of light neutrinos by using dark sector particles. The particles involved in the loop are associated with various phenomenologies. The mixing between the neutral component of $\Psi$ and $\chi$ gives rise to singlet-doublet dark matter (SDDM). The SDDM scenario has been widely investigated in the literature \cite{Cynolter:2015sua, Bhattacharya:2015qpa, Bhattacharya:2018fus, Bhattacharya:2017sml, Bhattacharya:2018cgx, Bhattacharya:2016rqj, Dutta:2020xwn, Borah:2021khc, Borah:2021rbx, Borah:2022zim, Borah:2023dhk, Paul:2024iie, Mahbubani:2005pt, DEramo:2007anh, Cohen:2011ec, Freitas:2015hsa, Calibbi:2015nha, Cheung:2013dua, Banerjee:2016hsk, DuttaBanik:2018emv, Horiuchi:2016tqw, Restrepo:2015ura, Abe:2017glm, Konar:2020wvl, Konar:2020vuu, Calibbi:2018fqf, Ghosh:2021wrk, Das:2023owa, Bhattacharya:2021ltd, Enberg:2007rp,Paul:2024prs,Paul:2025spm}. Recently, a comprehensive study of its entire parameter space has been performed, taking into account annihilation, co-annihilation, and conversion-driven processes for both Dirac \cite{Paul:2024prs} and Majorana \cite{Paul:2025spm} cases. An additional important feature of this set-up is the presence of new couplings between the singlet scalars and the SM Higgs field. While the electroweak phase transition (EWPT) in the SM is of second order, the inclusion of $\phi-H$ interaction can modify the dynamics of the Higgs potential, rendering the EWPT to be first order (FO) \cite{Curtin:2014jma}, even though $\phi$ does not get a vacuum expectation value (VEV). Such an electroweak first-order phase transition (EWFOPT) produces a stochastic gravitational wave (GW) background (see some pedagogical reviews \cite{Athron:2023xlk,Hindmarsh:2020hop}), which may fall within the sensitivity range of future GW observatories such as BBO \cite{Yunes:2008tw}, DECIGO \cite{Adelberger:2005bt}, and $\mu$ARES \cite{Sesana:2019vho}.

We note that all the new phenomena to be explained via the loop particles are strongly correlated with each other and therefore give rise to a new set of constraints on the parameter space. For example, if the $\phi-H$ interaction is not used to explain the FOPT, then the $\phi$ mass can be arbitrary while explaining the neutrino mass. However, from the FOPT, we observed that $\phi$ mass has to be less than a TeV while keeping the $\phi-H$ coupling within the perturbative limit. Similarly, the $\phi L\Psi$ interaction can drive a large lepton flavor violation (LFV). Therefore, the $\phi L\Psi$ coupling is not only constrained by the neutrino mass, but also by LFV. The upper bound on the $\phi L\Psi$ coupling also affects the singlet-doublet mixing, which is a crucial parameter for the realization of the DM relic abundance, which is an indirect requirement of satisfying the neutrino oscillation data. Thus, all these constraints collectively give rise to a parameter space, which can be verified in different terrestrial experiments.

The paper is organized as follows. In section \ref{sec:model}, we discuss the model in detail. Constraints from the leptonic sector are discussed in section \ref{sec:lepSect}. One-loop neutrino mass is discussed in subsection \ref{sbsec:numass}. The constraints from muon anomalous magnetic moment and charged lepton flavor violation are discussed in subsections \ref{sbsec:muong-2} and \ref{sbsec:lfv}, respectively. We discuss the DM phenomenology and its detection prospects in section \ref{sec:dmpheno}. Gravitational wave from the electroweak first-order phase transition is discussed in section \ref{sec:gw}, followed by a combined analysis of DM and GW parameter space in section \ref{sec:resultand discussion}. We finally conclude in section \ref{sec:concl}.

%%%%%%%%%%%%%%%%%%%%%%%%%%%%%%%%%%%%%%%%%%%%%%%%%%%%%%%%%%%%%%%%%%%%%%%%%%%%%%%%%%%%%%%%%%
\section{Model description}
\label{sec:model}
%%%%%%%%%%%%%%%%%%%%%%%%%%%%%%%%%%%%%%%%%%%%%%%%%%%%%%%%%%%%%%%%%%%%%%%%%%%%%%%%%%%%%%%%%%
We extend the SM with a vector-like doublet fermion, $\Psi$ $\left(=(\psi^0~~\psi^-)^T\equiv(\psi_L^0+\psi_R^0~~~~\psi^-)^T\right)$, a Majorana singlet fermion $\chi$, and three generations of singlet scalars $\phi_{1,2,3}$. The inclusion of three copies of $\phi$ facilitates masses to three generations of sub-eV neutrinos. We impose an additional $\mathcal{Z}_2$ symmetry under which $\Psi,\chi$, and $\phi_{1,2,3}$ are odd, while all other particles are even. The charge assignments of the particles under the imposed symmetry are listed in the table~\ref{tab:tab1}.
\begin{table}[h]
		\small
		\begin{center}
			\begin{tabular}{||@{\hspace{0cm}}c@{\hspace{0cm}}|@{\hspace{0cm}}c@{\hspace{0cm}}|@{\hspace{0cm}}c@{\hspace{0cm}}|@{\hspace{0cm}}c@{\hspace{0cm}}||}
				\hline
				\hline
				\begin{tabular}{c}
                {\bf ~~~~Symmetry~~~~}\\
				%{\bf ~~~~ Gauge~~~~}\\
					{\bf ~~~~Group~~~~}\\ 
					\hline
					
					$SU(2)_{L}$\\ 
					\hline
					$U(1)_{Y}$\\ 
					\hline
					$\mathcal{Z}_2$\\ 
				\end{tabular}
				&
				&
				\begin{tabular}{c|c|c}
					\multicolumn{3}{c}{\bf Fermion Fields}\\
					\hline
					~~~$L$~~~& ~~~$\Psi$~~~ & ~~~$\chi$~~~ \\
					\hline
					$2$&$2$&$1$\\
					\hline
					$-1$&$-1$&$0$\\
					\hline
					$+$&$-$&$-$\\
				\end{tabular}
				&
				\begin{tabular}{c|c|c}
					\multicolumn{2}{c}{\bf Scalar Field}\\
					\hline
					~~~$H$~~~& ~~~$\phi_{1,2,3}$~~~\\
					\hline
					$2$&$1$\\
					\hline
					$1$&$0$\\
                    \hline
					$+$&$-$\\
				\end{tabular}\\
				\hline
				\hline
			\end{tabular}
			\caption{Charge assignment of the fields under $SU(2)_L\otimes U(1)_Y\otimes\mathcal{Z}_2$ symmetry.}
			\label{tab:tab1}
		\end{center}    
	\end{table} 
\noindent
We assume the mass hierarchy of the particles running in the loop for neutrino mass generation as $M_{\chi}< M_{\Psi}, M_{\phi_i}$. We further assume $M_{\phi_1}<M_{\phi_2}<M_{\phi_3}$.
The relevant terms in the Lagrangian are given as,
\begin{eqnarray}
\mathcal{L}&=&i\bar{\Psi}\gamma^\mu D_\mu\Psi+i\bar{\chi}\gamma^\mu\partial_\mu\chi -M_{\Psi}\bar{\Psi}\Psi-\frac{1}{2}M_{\chi}\overline{\chi^C}\chi-\frac{y_{\chi}}{\sqrt{2}}\bar{\Psi}\tilde{H}(\chi+\chi^C)-y_{i\alpha}\bar{L}_{\alpha}\Psi\phi_i \nonumber\\&&- V(H,\phi_{1,2,3}) + {\rm h.c.}, \label{eq:lagrangian}
\end{eqnarray}
where $\alpha\in[e,\mu,\tau]$, and $i(=1,2,3)$ denotes the generation of singlet scalar $\phi$. The most general scalar potential is given as,
\begin{eqnarray}
    V(H,\phi_{1,2,3})&=&-\mu_h^2H^\dagger H+\lambda_h (H^\dagger H)^2+\mu_{1}^2\phi_1^2+\lambda_{1}\phi_{1}^4+\mu_{2}^2\phi_2^2+\lambda_{2}\phi_{2}^4\nonumber+\mu_{3}^2\phi_3^2+\lambda_{3}\phi_{3}^4\\&&+\lambda_{h1}H^\dagger H\phi_1^2+\lambda_{h2}H^\dagger H\phi_2^2+\lambda_{h3}H^\dagger H\phi_3^2+\lambda_{12}\phi_1^2\phi_2^2+\lambda_{13}\phi_1^2\phi_3^2+\lambda_{23}\phi_2^2\phi_3^2\nonumber\\&&+\lambda_{h12}H^\dagger H\phi_1\phi_2+\lambda_{h13}H^\dagger H\phi_1\phi_3+\lambda_{h23}H^\dagger H\phi_2\phi_3.
    \label{v0}
\end{eqnarray}
As the SM Higgs obtains a VEV, $v$, after EWSB, it induces a mixing between the singlet and neutral component of the doublet fermion through the $\bar{\Psi} \tilde{H} \chi$ coupling. The neutral fermion mass matrix then can be written in the basis $((\psi^0_R)^c,\psi^0_L,(\chi)^c)^T$ as,
\begin{equation}
\begin{pmatrix}
	0 & M_\Psi & \frac{m_D}{\sqrt{2}}\\
	M_\Psi&0&  \frac{m_D}{\sqrt{2}}\\
	 \frac{m_D}{\sqrt{2}} &  \frac{m_D}{\sqrt{2}}& M_\chi
\end{pmatrix},
\end{equation}
where $m_D=y_\chi v/\sqrt{2}$.
The mass matrix can be diagonalized with a unitary matrix of the form $U(\theta)=U_{13}(\theta_{13}=\theta).U_{23}(\theta_{23}=0).U_{12}(\theta_{12}=\frac{\pi}{4})$.
The three neutral states mix and give three Majorana states as $\chi_i=(\chi_{iL}+\chi_{iL}^C)/\sqrt{2}$, where
\begin{eqnarray}
\chi_{1L}&=&\frac{\cos\theta}{\sqrt{2}}(\psi^0_L+(\psi^0_R)^C)+\sin\theta \chi^C,\nonumber\\
\chi_{2L}&=&\frac{i}{\sqrt{2}}(\psi^0_L-(\psi^0_R)^C),\nonumber\\
\chi_{3L}&=&-\frac{\sin\theta}{\sqrt{2}}(\psi^0_L+(\psi^0_R)^C)+\cos\theta \chi^C.
\end{eqnarray}
The corresponding mass eigenvalues are
\begin{eqnarray}
M_{\chi_1}&=&M_\Psi\cos^2\theta+M_\chi\sin^2\theta+m_D\sin2\theta,\nonumber\\M_{\chi_2}&=&M_\Psi,\nonumber\\
M_{\chi_3}&=&M_\Psi\sin^2\theta+M_\chi\cos^2\theta-m_D\sin2\theta,
\label{fermion masses}
\end{eqnarray}
where the mixing angle is given as
\begin{eqnarray}
	\tan2\theta=\frac{2m_D}{M_\Psi-M_\chi}.
\end{eqnarray}
Here we identify the $\chi_3$ to be DM candidate. The Yukawa coupling can be expressed as,
\begin{eqnarray}\label{eq:theta_deltaM}
y_\chi=\frac{{\Delta}M\sin2\theta}{\sqrt{2}v},
\end{eqnarray}
where ${\Delta}M$ is the mass splitting between DM and the next heavy neutral fermion state. The important parameters in the DM phenomenology are $\{M_{\chi_3}\equiv M_{{\rm DM}},{\Delta}M=M_{\chi_1}-M_{\chi_3}\approx M_{\chi_2}-M_{\chi_3},\sin\theta,{\Delta}M'=M_{\phi_1}-M_{\chi_3},\lambda_{h1}\}$.

%%%%%%%%%%%%%%%%%%%%%%%%%%%%%%%%%%%%%%%%%%%%%%%%%%%%%%%%%%%%%%%%%%%%%%%%%%%%%%%%%%%%%%%%%%
\section{Lepton Sector Constraints}
\label{sec:lepSect}
%%%%%%%%%%%%%%%%%%%%%%%%%%%%%%%%%%%%%%%%%%%%%%%%%%%%%%%%%%%%%%%%%%%%%%%%%%%%%%%%%%%%%%%%%%
It is evident from the discussion in the previous section that the leptonic sector in the model provides interesting phenomenologies. In this section, we lay down the relevant effects from the leptonic sector.  
%%%%%%%%%%%%%%%%%%%%%%%%%%%%%%%%%%%%%%%%%%%%%%%%%%%%%%%%%%%%%%%%%%%%%%%%%%%%%%%%%%%%%%%%%%
\subsection{Neutrino mass}
\label{sbsec:numass}
%%%%%%%%%%%%%%%%%%%%%%%%%%%%%%%%%%%%%%%%%%%%%%%%%%%%%%%%%%%%%%%%%%%%%%%%%%%%%%%%%%%%%%%%%%
\begin{figure}[H]
    \centering
    \includegraphics[scale=0.4]{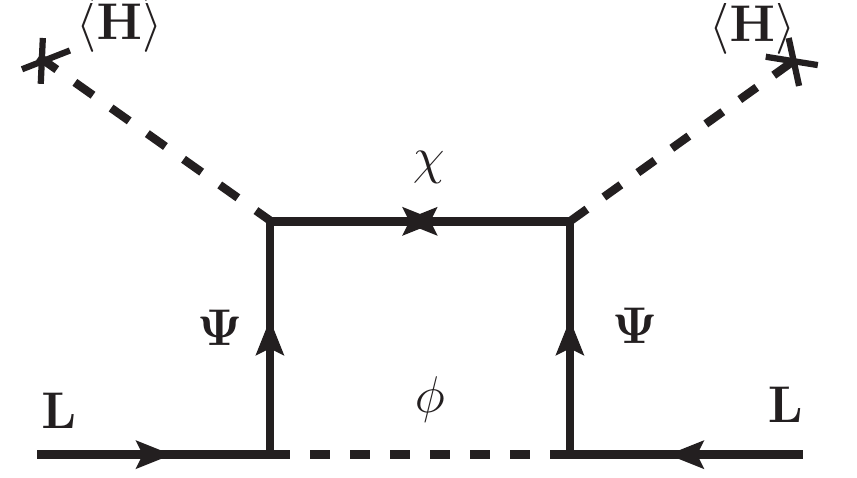}
    \caption{One loop realization of Majorana neutrino mass using dark sector particles in the loop.}
    \label{fig:numass}
\end{figure}
In this model, the lepton number-violating Weinberg operator \cite{Weinberg:1979sa}, $LLHH$, can be realized at one-loop level through the diagram presented in Fig. \ref{fig:numass} \cite{Fraser:2014yha,Borah:2022zim,Konar:2020wvl,Mohapatra:2022ngo}, where all the dark sector particles run in the loop. The radiative neutrino mass, in the effective theory, is given by (see Appendix \ref{app:numass_matrix} for details.)
\begin{eqnarray}\label{eq:numass}
\left(m_{\nu}\right)_{\alpha\beta}&=&\sum_{i=1}^3\frac{y_{\chi }^2y_{\alpha{i}}y_{{i}\beta}v^2}{8\pi^2} M_\chi~\left[\frac{M_\chi^4}{(M_\chi^2-M_{\phi_i}^2)(M_\chi^2-M_\Psi^2)^2}\log\left[\frac{M_\chi^2}{M_\Psi^2}\right]\right.\nonumber\\
    &&~~~\left.+\frac{M_\Psi^2}{(M_\chi^2-M_\Psi^2)(M_{\phi_i}^2-M_\Psi^2)}-\frac{M_{\phi_i}^4}{(M_\chi^2-M_{\phi_i}^2)(M_{\phi_i}^2-M_\Psi^2)^2}\log\left[\frac{M_{\phi_i}^2}{M_\Psi^2}\right]\right],
\end{eqnarray}
where 
\begin{align}
    M_\chi&=\frac{1}{\cos2\theta}\left(M_{\chi_3}\cos^2\theta - M_{\chi_1}\sin^2\theta+m_D\sin2\theta\right),\nonumber\\
    M_\Psi&=\frac{1}{\cos2\theta}\left(M_{\chi_1}\cos^2\theta - M_{\chi_3}\sin^2\theta-m_D\sin2\theta\right).
\end{align}
The structure of the neutrino mass matrix can be expressed as,
\begin{equation}
    (m_\nu )_{\alpha\beta}=(y^T \mathcal{M} y)_{\alpha\beta},
\end{equation}
where $\mathcal{M}$ is a diagonal matrix whose $i$-th component is given by,
\begin{eqnarray}
    \mathcal{M}_i&=&\frac{y_{\chi }^2v^2}{8\pi^2} M_\chi~\left[\frac{M_\chi^4}{(M_\chi^2-M_{\phi_i}^2)(M_\chi^2-M_\Psi^2)^2}\log\left[\frac{M_\chi^2}{M_\Psi^2}\right]\right.\nonumber\\
    &&~~~~~~~\left.+\frac{M_\Psi^2}{(M_\chi^2-M_\Psi^2)(M_{\phi_i}^2-M_\Psi^2)}-\frac{M_{\phi_i}^4}{(M_\chi^2-M_{\phi_i}^2)(M_{\phi_i}^2-M_\Psi^2)^2}\log\left[\frac{M_{\phi_i}^2}{M_\Psi^2}\right]\right].
\end{eqnarray}
Using the Casas-Ibarra parameterization \cite{Casas:2001sr,Herrero-Garcia:2025aox}, the Yukawa coupling can be expressed as,
\begin{equation}
    y=\sqrt{\mathcal{M}^{-1}}R\sqrt{\hat{m}_\nu}U_{\rm PMNS},
\end{equation}
where $\hat{m}_\nu$ is the diagonal neutrino mass matrix and $U_{\rm PMNS}$ is the Pontecorvo-Maki-Nakagawa-Sakata matrix. The complex orthogonal matrix, $R$ can be parameterized as,
\begin{equation}
    R=\begin{pmatrix}
        1 & 0& 0\\
        0 &\cos{\alpha} & \sin{\alpha}\\
        0 & -\sin{\alpha}& \cos{\alpha}
    \end{pmatrix}
    \begin{pmatrix}
        \cos{\beta}& 0 & \sin{\beta}\\
        0 & 1 & 0\\
        -\sin{\beta} & 0 & \cos{\beta}
    \end{pmatrix}
    \begin{pmatrix}
        \cos{\gamma}&\sin{\gamma}&0\\
        -\sin{\gamma}&\cos{\gamma}&0\\
        0 & 0 & 1
    \end{pmatrix},
\end{equation}
where $\alpha,\beta$ and $\gamma$ represent complex angles. The real and imaginary parts of $\alpha,\beta$ and $\gamma$ are randomly varied within $[-\pi,\pi]$. We have used the values of neutrino oscillation parameters in the $3\sigma$ range from \cite{deSalas:2020pgw} for the rest of our analysis, as given in table \ref{tab:tab2}.
\begin{table}[h]
\begin{center}
		\resizebox{6cm}{!}{
			\begin{tblr}{
					colspec={|l|l|}%,
					% row{1}={font=\bfseries},
					% column{1}={font=\itshape}
				}
				\toprule Parameters & Values in $3\sigma$ range\\
				\toprule
				$\Delta m_{21}^2[10^{-5}\rm eV^2]$&6.94–8.14\\
                \hline
				$\Delta m_{31}^2[10^{-3}\rm eV^2]$&2.47–2.63\\
                \hline
				$\sin^2\theta_{12}$&0.271–0.369\\
                \hline
				$\sin^2\theta_{23}$&0.434–0.61\\
                \hline
				$\sin^2\theta_{13}$&0.02–0.02405\\
                \hline
				$\delta$&128\degree–359\degree\\
				\bottomrule
		\end{tblr}}
		\caption{The $3\sigma$ ranges of the neutrino oscillation parameters\cite{deSalas:2020pgw} for normal hierarchy spectrum of the neutrino mass.}
		\label{tab:tab2}
\end{center}
\end{table}

%%%%%%%%%%%%%%%%%%%%%%%%%%%%%%%%%%%%%%%%%%%%%%%%%%%%%%%%%%%%%%%%%%%%%%%%%%%%%%%%%%%%%%%%%%
\subsection{Muon anomalous magnetic moment}
\label{sbsec:muong-2}
\begin{figure}[h]
    \centering
    \includegraphics[scale=0.5]{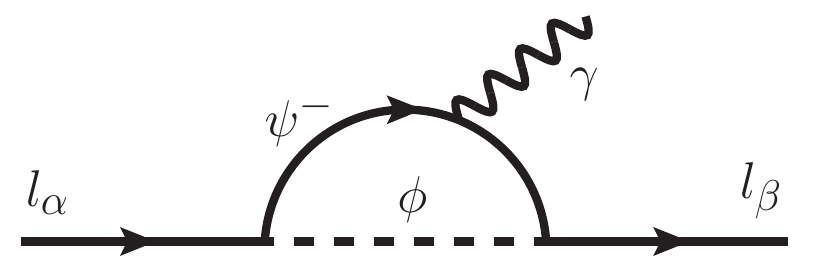}
    \caption{The Feynman diagram giving rise $(g-2)_\mu$ and charged lepton flavor violation.}
    \label{fig:lfv}
\end{figure}

%%%%%%%%%%%%%%%%%%%%%%%%%%%%%%%%%%%%%%%%%%%%%%%%%%%%%%%%%%%%%%%%%%%%%%%%%%%%%%%%%%%%%%%%%%
In our setup, the new positive contribution to the muon $(g-2)$ comes from the one-loop diagram with the charged doublet fermion $\psi^{-}$ and the singlet scalars $\phi_i$s in the loop as shown in Fig. \ref{fig:lfv}.
This contribution to $(g-2)$ can be estimated as \cite{Lindner:2016bgg,Athron:2025ets},
\begin{equation}\label{eq:10}
\Delta a_\mu=\sum_{i}\frac{y_{i\mu}^2}{8\pi^2}\frac{m^2_\mu}{M^2_{\phi_i}}\int^1_0~dx\frac{x^2(1-x+\frac{M_{\Psi}}{m_\mu})}{(1-x)(1-x\frac{m^2_\mu}{M^2_{\phi_i}})+x\frac{M^2_{\Psi}}{M^2_{\phi_i}}}
\end{equation}
The present value of the muon anomalous magnetic moment is reported to be $\Delta{a_{\mu}}=38(63)\times10^{-11}$ \cite{Muong-2:2025xyk,Aliberti:2025beg}. This indicates that, within the current uncertainties, there is no statistically significant tension between the experimental measurement and the SM prediction. Nevertheless, we use the upper limit on $\Delta{a_{\mu}}=101\times10^{-11}$ to constrain the model parameter space.

%%%%%%%%%%%%%%%%%%%%%%%%%%%%%%%%%%%%%%%%%%%%%%%%%%%%%%%%%%%%%%%%%%%%%%%%%%%%%%%%%%%%%%%%%%
\subsection{Charged lepton flavor violation}
\label{sbsec:lfv}
%%%%%%%%%%%%%%%%%%%%%%%%%%%%%%%%%%%%%%%%%%%%%%%%%%%%%%%%%%%%%%%%%%%%%%%%%%%%%%%%%%%%%%%%%%
The observation of neutrino oscillation has indicated that the lepton number is not an exact symmetry of the SM. This allows charged lepton flavor violating (cLFV) processes within the SM at the one-loop level, and they are suppressed by the tiny neutrino mass.

At this stage, any cLFV processes such as $\mu\rightarrow e\gamma$ may act as a smoking gun signature for BSM physics. In our setup, the singlet scalars and the doublet fermion allow cLFV\footnote{We note that the channel $\mu \rightarrow e\gamma$ imposes the most stringent constraint on the parameter space. In contrast, the current experimental bounds on the channels associated with $\tau$ decays into $e$ or $\mu$ are comparatively weak \cite{MEGII:2023ltw}. Therefore, we use only the $\mu \rightarrow e\gamma$ process to constrain our model parameters.} interactions via loop mediated processes shown in Fig. \ref{fig:lfv}.
The branching ratio for $\mu\rightarrow e\gamma$ is given by \cite{Lindner:2016bgg},
\begin{align}
     {\rm Br} (\mu\rightarrow e\gamma) &= \sum_{i} \frac{3 \alpha y_{ie}^2 y_{i\mu}^2}{16 \pi G_F^2}\nonumber \\
     &\times \left(\int_0^1 \int_0^{1-x} dy~ dx\frac{x\left(y+(1-x-y)\frac{m_e}{m_\mu}\right)+(1-x)\frac{M_{\Psi}}{m_\mu}}{-xy m_\mu^2-x(1-x-y)m_e^2+x M_{\phi_i}^2+(1-x)M_{\Psi}^2}\right)^2
\end{align}
\begin{figure}[htbp]
    \centering
    \includegraphics[scale=0.5]{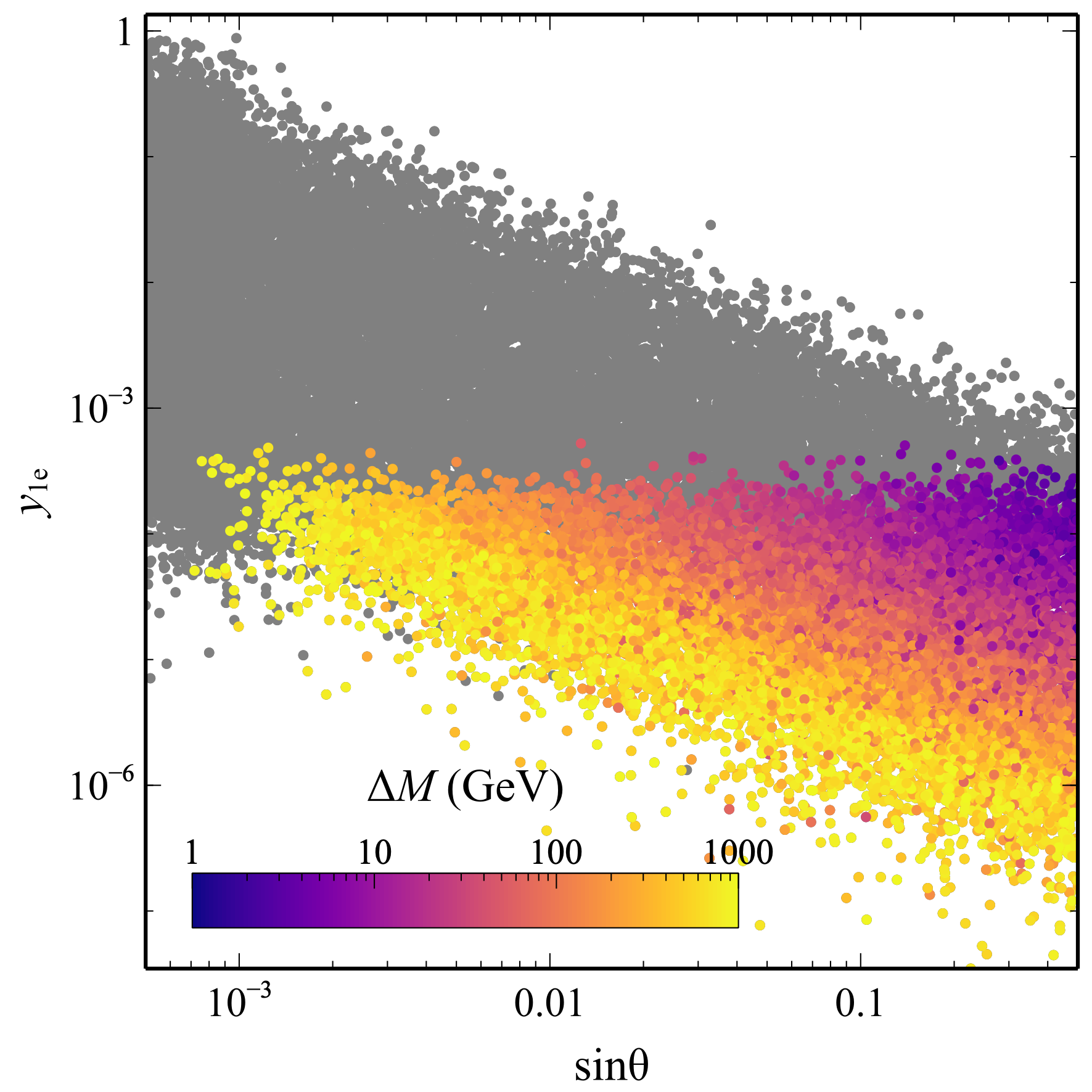}
    \caption{The allowed parameter space from the neutrino mass is shown with the gray points in the plane of $y_{1e}$ vs $\sin\theta$. The colored points are consistent with the charged lepton flavor violation constraint. The color code represents the SD mass splitting, $\Delta M$.}
    \label{fig:lfvparams}
\end{figure}
The most recent constraint from the MEG-II collaboration sets an upper limit of ${\rm Br}(\mu\rightarrow e\gamma)<3.1\times10^{-13}$ at 90\% C.L. \cite{MEGII:2023ltw}. We impose this bound on our parameter space to ensure consistency with the desired phenomenology. In Fig. \ref{fig:lfvparams}, we have shown SD model parameters in the plane of $y_{1e}$ and $\sin\theta$, where the color code represents SD mass splitting, $\Delta M$. The gray colored points satisfy the neutrino mass constraints, while only the colored points with typically $y_{1e}\lesssim 10^{-3}$ are allowed by the combined constraints from cLFV, $(g-2)_\mu$, and neutrino mass. From Fig. \ref{fig:lfvparams}, we see that all values of $\sin{\theta}\gtrsim\mathcal{O}(10^{-3})$ are allowed by both neutrino mass and cLFV. The neutrino mass, as given in Eq. (\ref{eq:numass}), is proportional to the Yukawa couplings ($y_{i\alpha}$) and $y_{\chi}$. From Eq. (\ref{eq:numass}) and Eq. (\ref{eq:theta_deltaM}), we see that, for a fixed $\Delta M$,  a decrease in $y_{1e}$ results an increase in $\sin\theta$ (or  $y_\chi$) to satisfy neutrino oscillation data. Similarly, for a fixed $\sin\theta$, a decrease in $\Delta M$ leads to a decrease in $y_\chi$. This implies a larger $y_{1e}$ is required to satisfy neutrino oscillation data.

%%%%%%%%%%%%%%%%%%%%%%%%%%%%%%%%%%%%%%%%%%%%%%%%%%%%%%%%%%%%%%%%%%%%%%%%%%%%%%%%%%%%%%%%%%
\section{Dark matter phenomenology}
\label{sec:dmpheno}
%%%%%%%%%%%%%%%%%%%%%%%%%%%%%%%%%%%%%%%%%%%%%%%%%%%%%%%%%%%%%%%%%%%%%%%%%%%%%%%%%%%%%%%%%%
\subsection{Relic density}

In this model, the thermal relic of SD Majorana DM is realized via the freeze-out of various processes such as annihilation, co-annihilation, co-scattering, decay, and inverse decay depending on the values of SD mixing angle ($\sin\theta$) and the mass splitting between DM and ``next to lightest stable particle'' (NLSP). 
To incorporate all the contributions, we categorize the particles into three sectors: sector 1, which includes the singlet DM, $\chi_3$; sector 2, which includes $\chi_1,\chi_2,\psi^-$, and $\phi_{1,2,3}$; and sector 0, which includes all the SM particles. Now we define the co-moving number density of the sector 1 and sector 2 particles as $Y_1=n_{\chi_3}/s$, and $Y_2=(n_{\chi_1}+n_{\chi_2}+n_{\psi^-}+n_{\phi_1}+n_{\phi_2}+n_{\phi_3})/s$, where $n_i$ is the number density of $i$-th species, and  $s=2\pi^2 g_{*s}T^3/45$ is the entropy density. The evolution of DM and other dark sector particles are governed by the following Boltzmann Equations (BEs) \cite{Paul:2025spm},
\begin{eqnarray}
		\frac{dY_1}{dT} &=&   \frac{1}{3\mathcal{H}}\frac{ds}{dT} \left[    \langle \sigma_{1100} v \rangle ( Y_1^2 - {Y_1^{\rm eq}}^2) +    \langle \sigma_{1122} v \rangle \left( Y_1^2 - Y_2^2  \frac{{Y_1^{\rm eq}}^2}{{Y_2^{\rm eq}}^2}\right)  + \langle \sigma_{1200} v \rangle ( Y_1 Y_2 - Y_1^{\rm eq}Y_2^{\rm eq})\right. \nonumber\\
		&&+\left.  \langle \sigma_{1222} v \rangle \left( Y_1 Y_2 - Y_2^2   \frac{Y_1^{\rm eq}}{Y_2^{\rm eq}} \right) -\langle \sigma_{1211} v \rangle \left( Y_1 Y_2 - Y_1^2   \frac{Y_2^{\rm eq}}{Y_1^{\rm eq}} \right)
		-\frac{ \Gamma_{2\rightarrow 1}}{s}\left( Y_2 -Y_1 \frac{Y_2^{\rm eq}}{Y_1^{\rm eq}}  \right) \right]\nonumber\\        
		\label{eq:Y1}
	\end{eqnarray}
	\begin{eqnarray}
		\frac{dY_2}{dT} &=&   \frac{1}{3\mathcal{H}}\frac{ds}{dT}\left[    \langle \sigma_{2200} v \rangle ( Y_2^2 - {Y_2^{\rm eq}}^2) -    \langle \sigma_{1122} v \rangle \left( Y_1^2 - Y_2^2  \frac{{Y_1^{\rm eq}}^2}{{Y_2^{\rm eq}}^2}\right) +  \langle \sigma_{1200} v \rangle ( Y_1 Y_2 - Y_1^{\rm eq}Y_2^{\rm eq}) \right. \nonumber \\
		&&- \left. \langle \sigma_{1222} v \rangle \left( Y_1 Y_2 - Y_2^2   \frac{Y_1^{\rm eq}}{Y_2^{\rm eq}} \right)
		+\langle \sigma_{1211} v \rangle \left( Y_1 Y_2 - Y_1^2   \frac{Y_2^{\rm eq}}{Y_1^{\rm eq}} \right)  + \frac{ \Gamma_{2\rightarrow 1}}{s}\left( Y_2 -Y_1 \frac{Y_2^{\rm eq}}{Y_1^{\rm eq}}  \right)                \right],\nonumber\\
		\label{eq:Y2}
\end{eqnarray}
where $Y_i^{\rm eq}\left(=n_i^{\rm eq}/s\right)$ are the equilibrium abundances, $\mathcal{H}=1.66\sqrt{g_*}T^2/M_{\rm Pl}$ is the  Hubble parameter with $M_{\rm Pl}=1.22\times10^{19}$ GeV being the Planck mass, and $\langle \sigma_{\alpha\beta\gamma\delta} v\rangle$ are the thermally averaged cross-sections for processes involving the annihilation of particles of sectors $\alpha\beta\rightarrow \gamma\delta$, which is given by \cite{Gondolo:1990dk,Alguero:2022inz},
    \begin{eqnarray}
      \langle \sigma_{\alpha\beta\gamma\delta} v\rangle=\frac{T}{8m_{\alpha}^2m_{\beta}^2K_{2}(\frac{m_\alpha}{T})K_{2}(\frac{m_\beta}{T})}\int_{(m_\alpha+m_\beta)^2}^\infty\sigma_{\alpha\beta\rightarrow\gamma\delta}(s)\big(s-(m_\alpha+m_\beta)^2\big)\sqrt{s}K_1\bigg(\frac{\sqrt{s}}{T}\bigg)ds,\nonumber\\ 
    \end{eqnarray}
and $\Gamma_{2\rightarrow 1}$ in Eq. (\ref{eq:Y1}) is the conversion term, which includes both the interaction rate of the co-scattering process as well as the decay and is given  as   
\begin{eqnarray}\label{eq:gamma21}
\Gamma_{2\rightarrow1}=\Gamma_{2\rightarrow\chi_3,\rm SM}\frac{K_1(M_{\Psi}/T)}{K_2(M_{\Psi}/T)}+\langle \sigma_{2010} v \rangle n^{\rm eq}_{\rm SM},
\end{eqnarray}
where $\Gamma_{2\rightarrow\chi_3,\rm SM}$ includes all the two body and three body decays of sector 2 particles to sector 1 particle, $\langle\sigma_{2010}v\rangle$ denotes the thermally averaged cross-sections of the co-scattering processes. 
The relevant parameters in the DM relic calculation are,
\begin{align*}
\{M_{\rm DM} &\equiv M_{\chi_3},\sin\theta, \Delta{M}=M_{\chi_1}-M_{\rm DM}, \Delta{M^\prime}=M_{\phi_1}-M_{\rm DM},
\lambda_{h1}\}.
\end{align*}

We compute the relic of DM using the {\tt micrOMEGAs} package \cite{Alguero:2023zol}. 
\begin{figure}[h]
    \centering
    \includegraphics[scale=0.4]{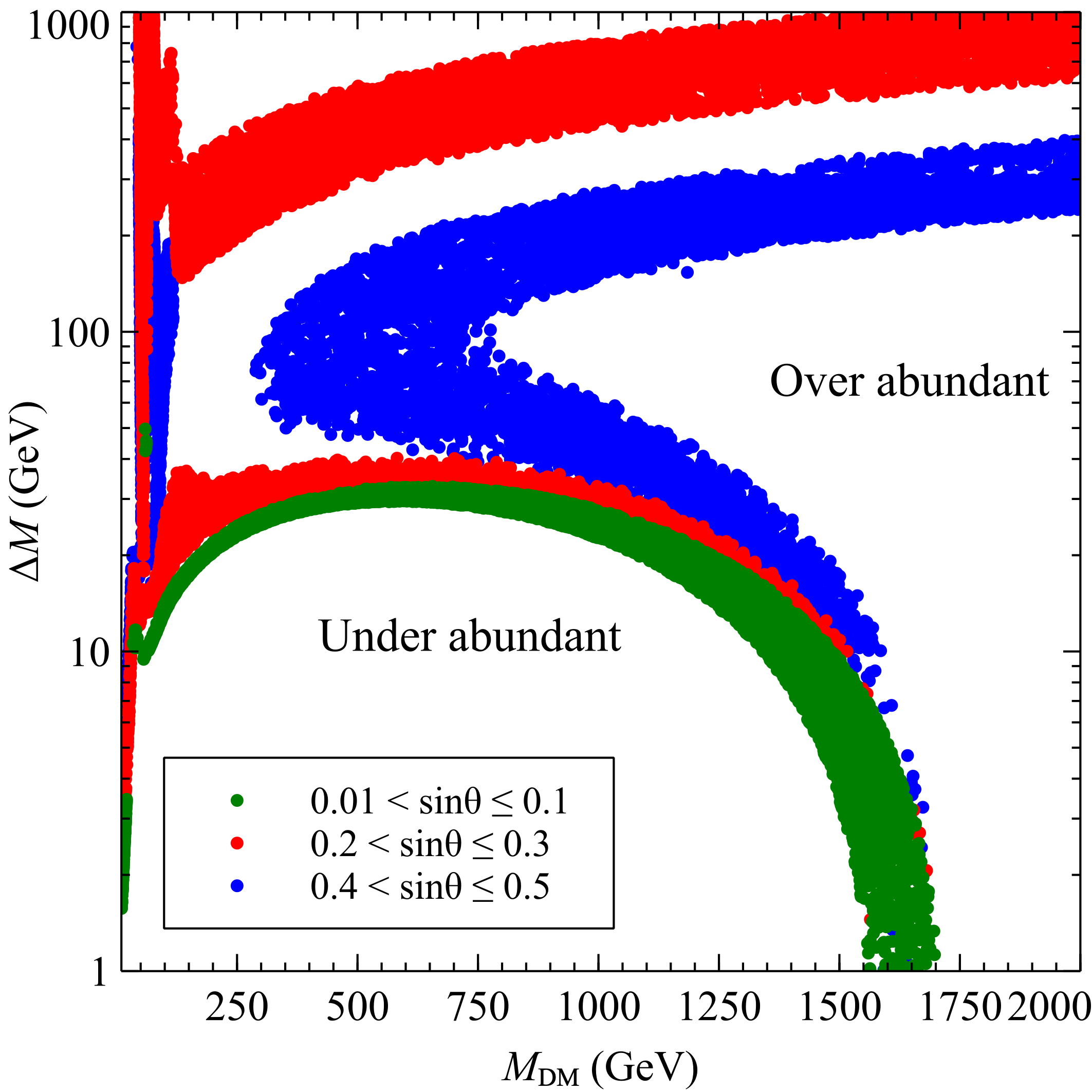}
    \caption{Correct DM relic parameter space in the plane of $\Delta{M}$ vs $M_{\rm DM}$ considering $\Delta M\ll\Delta{M}^\prime$. This plot does not yet include direct detection constraint.}
    \label{fig:delmVSmdm}
\end{figure}
\begin{figure}[h]
    \centering
    \includegraphics[scale=0.5]{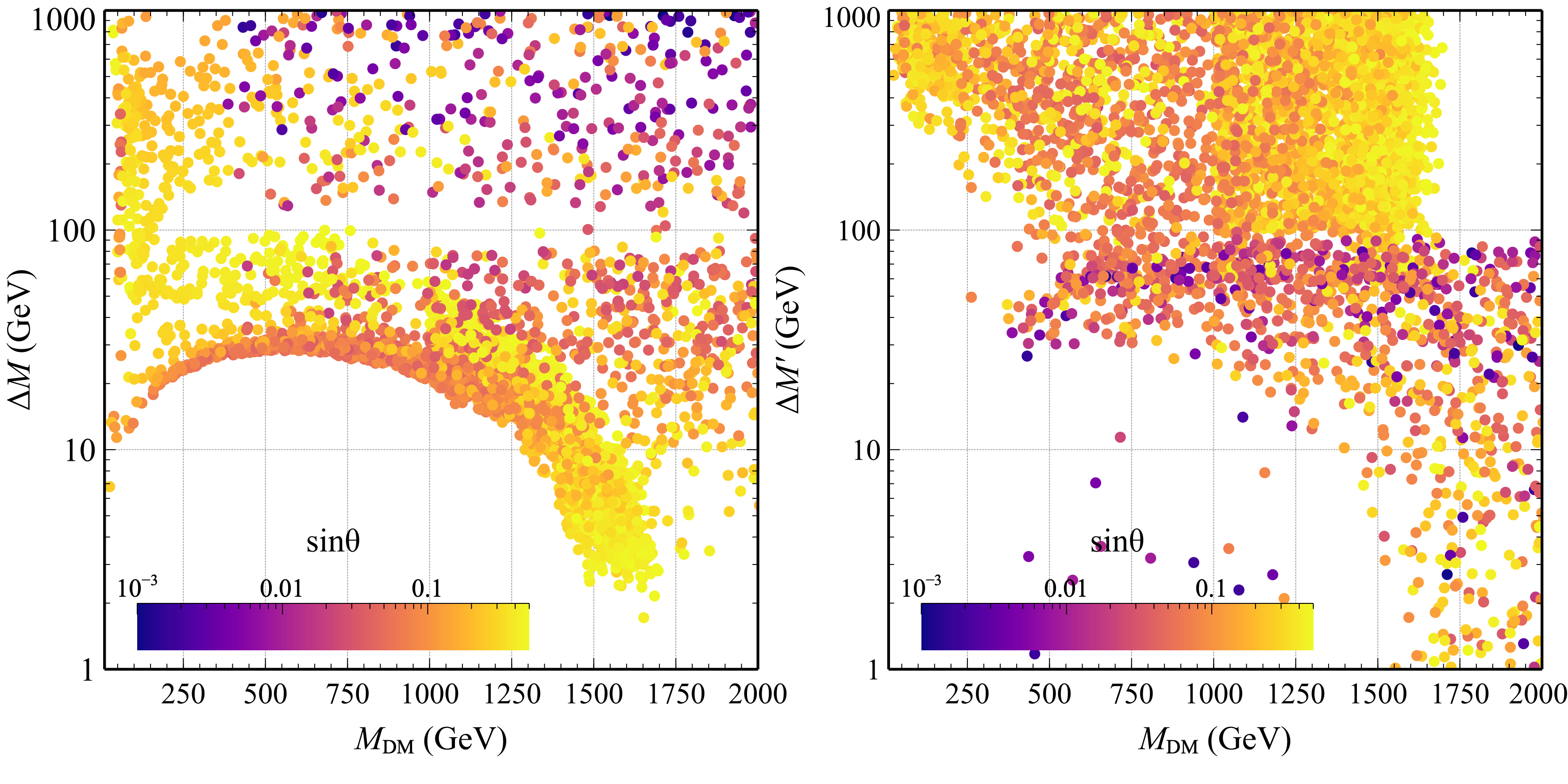}
    \caption{\textit{Left:} DM parameter space satisfying correct relic in the plane of $\Delta{M}-M_{\rm DM}$. \textit{Right:} the same points are shown in the plane of $\Delta{M^\prime}-M_{\rm DM}$. The SD mixing angle is shown in the color code. All these points satisfy the constraints from neutrino mass, $(g-2)_\mu$, and cLFV. These plots do not yet include direct detection constraint.}
    \label{fig:dmVSmdm_CR}
\end{figure}
If $\Delta{M}\ll\Delta{M}^\prime$ ($\Delta{M}\gg\Delta{M}^\prime$), then $\chi_1,\chi_2,\psi^-$ ($\phi_1$) will play significant role in the final relic. On the other hand, if $\Delta{M}\simeq\Delta{M}^\prime$, all the sector 2 particles will contribute to the relic density. We also note that for $\sin\theta\gtrsim\mathcal{O}(10^{-2})$ only annihilation and co-annihilation are important for the calculation of the relic density, while, for small $\sin\theta$ (typically $\sin\theta\lesssim\mathcal{O}(10^{-3})$) conversion-driven processes play an significant role along with co-annihilation depending on the values of $\Delta{M}$ and $M_{\rm DM}$. We observe from Fig. \ref{fig:lfvparams} that for small $\sin\theta$ (typically $\lesssim 10^{-2}$), cLFV restricts $\Delta M\gtrsim 50$ GeV. Moreover, we see that all values of $\sin\theta\lesssim\mathcal{O}(10^{-3})$ are disallowed by cLFV processes. Therefore, the contribution from the conversion-driven processes to the relic density is almost negligible in our model \cite{Paul:2024prs,Paul:2025spm}. However, for completeness, we have solved the BEs considering annihilation, co-annihilation, and conversion-driven processes as given in Eqs. (\ref{eq:Y1}) and (\ref{eq:Y2}).

In the Fig.~\ref{fig:delmVSmdm}, we present the parameter space that satisfy the observed relic density in the $\Delta M$–$M_{\rm DM}$ plane, focusing on the region where co-annihilation between the dark matter (DM) and $\phi_1$ is subdominant (i.e., for $\Delta M' \gg \Delta{M}$). As we go from left to right, the DM mass increases, the annihilation cross-section decreases, and the relic density increases. The relic can be brought to the correct ballpark by decreasing the mass splitting, $\Delta{M}$. This is clearly visible in the Fig. \ref{fig:delmVSmdm}. For larger values of $\Delta{M}$, co-annihilation becomes negligible, and Higgs-mediated annihilation processes become relevant in deciding the relic of DM. The Yukawa coupling is proportional to the $\Delta{M}\sin2\theta$. For a given $\sin\theta$, larger $\Delta{M}$ leads to stronger coupling and hence larger annihilation cross-section. This results in a smaller relic density. To get the correct relic abundance, the DM mass must increase to compensate for this effect. For a typical $\sin\theta$, say $0.4 < \sin{\theta} < 0.5$ (blue colored points), the region to the right corresponds to an over-abundant region, while the region to the left represents an under-abundant region.
Further reducing $\sin{\theta}$, say $0.2<\sin\theta<0.3$ (red colored points), leads to an overall decrease in annihilation and co-annihilation cross sections, thereby converting the earlier under-abundant region (left to blue colored points) to the relic density satisfying regime. 

Now we include the effect of $\phi_1$ in relic determination by allowing $\Delta M'<100$ GeV values. In the left panel of Fig. \ref{fig:dmVSmdm_CR}, we have presented the points satisfying the relic density on the plane of $\Delta M$ vs. $M_{\rm DM}$ and the same points are also shown in $\Delta M^\prime$ vs. $M_{\rm DM}$ plane in the right panel of Fig. \ref{fig:dmVSmdm_CR}. The color code represents $\sin\theta$. 
In the present framework, we assume $\phi_{2,3}$ to be much heavier than $\phi_1$ to reduce the computational complexity. Therefore, the only new degrees of freedom (DoF) in comparison to earlier discussion is $\phi_1$. This opens up new annihilation and co-annihilation channels, in addition to the standard DM self-annihilation and singlet–doublet co-annihilation processes through $y_{1\alpha}\phi_1 L_\alpha\Psi$ coupling. These processes are given in Appendix \ref{app:feyndiags}. Note that the effective coupling strength between  DM and $\phi_1$ co-annihilation is $\propto y_{1\alpha}\sin\theta $. The inclusion of these additional channels enables the model to reproduce the correct DM relic abundance even within regions that would otherwise be over-abundant. In particular, the white region, right to the blue colored points in Fig. \ref{fig:delmVSmdm}, is now satisfying the correct relic density. This feature is evident from the \textit{left} panel of Fig.~\ref{fig:dmVSmdm_CR}.  From the right panel of Fig. \ref{fig:dmVSmdm_CR}, we see that all the new points representing the over-abundant region in the absence of $\phi_1$ correspond to $\Delta{M}^\prime<100$ GeV. It is worth mentioning again that the points in Fig. \ref{fig:delmVSmdm} and \ref{fig:dmVSmdm_CR} satisfy the constraints from neutrino mass, $(g-2)_\mu$, cLFV, and correct DM relic. We discuss the direct detection constraints in the following section.

%%%%%%%%%%%%%%%%%%%%%%%%%%%%%%%%%%%%%%%%%%%%%%%%%%%%%%%%%%%%%%%%%%%%%%%%%%%%%%%%%%%%%%%%%%
\subsection{Direct Detection}
\label{sec:dd}
%%%%%%%%%%%%%%%%%%%%%%%%%%%%%%%%%%%%%%%%%%%%%%%%%%%%%%%%%%%%%%%%%%%%%%%%%%%%%%%%%%%%%%%%%%
\begin{figure}[H]
    \centering
    \includegraphics[scale=1]{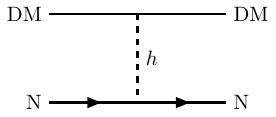}
    \caption{Feynman diagram for spin-independent DM-nucleon scattering cross-section}
    \label{fig:ddFD}
\end{figure}
The cross section per nucleon for the spin-independent (SI) DM-nucleon interaction mediated by SM Higgs, as shown in Fig. \ref{fig:ddFD}, is given by,
\begin{equation}
\label{da}
    \sigma_{\rm SI} = \frac{1}{\pi A^2}\mu^2_r|\mathcal{M}|^2,
\end{equation}
where A is the mass number of the target nucleus, $\mu_r$ is the reduced mass of the DM-nucleon system, and ${\mathcal M}$ 
is the amplitude for the DM-nucleon interaction, which can be written as
\begin{equation}
\label{dd}
    \mathcal{M}=\Big[Z f_p +(A-Z)f_n\Big],
\end{equation}
where $f_{p}$ and $f_{n}$ denote effective interaction strengths of DM with proton and neutron of the nuclei used for the experiment, with $Z$ being the atomic number. The effective interaction strength can then further be decomposed in terms of interaction with the partons as
\begin{equation}
\label{dda}
    f_{p,n}=\sum_{q=u,d,s}f^{p,n}_{Tq}\alpha_{q}\frac{m_{(p,n)}}{m_q} + \frac{2}{27}f^{p,n}_{TG}\sum_{q=c,b,t}\alpha_q \frac{m_{(p,n)}}{m_{q}};
\end{equation}  
with 
\begin{equation}
\label{dda2}
\alpha_q =\frac{y_\chi \sin2\theta}{M^2_h}\frac{m_q}{v}=\frac{~\Delta M ~\sin^22\theta m_q}{\sqrt{2}v^2 M^2_h}; 
\end{equation}
coming from DM interaction with SM via Higgs portal coupling. Further, in Eq.~(\ref{dda}), the different coupling strengths between DM and light quarks are given as $f^p_{Tu} = 0.020 \pm 0.004, f^p_{Td} = 0.026 \pm 0.005, f^p_{Ts} = 0.014 \pm 0.062$, $f^n_{Tu} = 0.020 \pm 0.004, 
f^n_{Td} = 0.036 \pm 0.005, f^n_{Ts} = 0.118 \pm 0.062$ \cite{Bertone:2004pz,Alarcon:2012nr}. The coupling of DM with the gluons in target nuclei is parameterized by
\begin{equation*}
f^{(p,n)}_{TG} = 1- \sum_{q=u,d,s}f^{p,n}_{Tq}.
\end{equation*}
Using Eqs.~(\ref{da}), (\ref{dd}), (\ref{dda}) and (\ref{dda2}), the spin-independent DM-nucleon cross-section is given by
\begin{equation}
\label{ddaf}
 \sigma_{\rm SI} = \frac{4 \mu^2_r}{\pi A^2}\frac{y_\chi^2 \sin^2 2\theta}{M^4_h}\Big[\frac{m_p}{v}\Big(f^{p}_{Tu} + f^{p}_{Td} + f^{p}_{Ts} + \frac{2}{9}f^{p}_{TG}+\frac{m_n}{v}\Big(f^{n}_{Tu} + f^{n}_{Td} + f^{n}_{Ts} + \frac{2}{9}f^{n}_{TG}\Big)\Big]^2.
\end{equation}

\begin{figure}[t]
    \centering
    \includegraphics[scale=0.35]{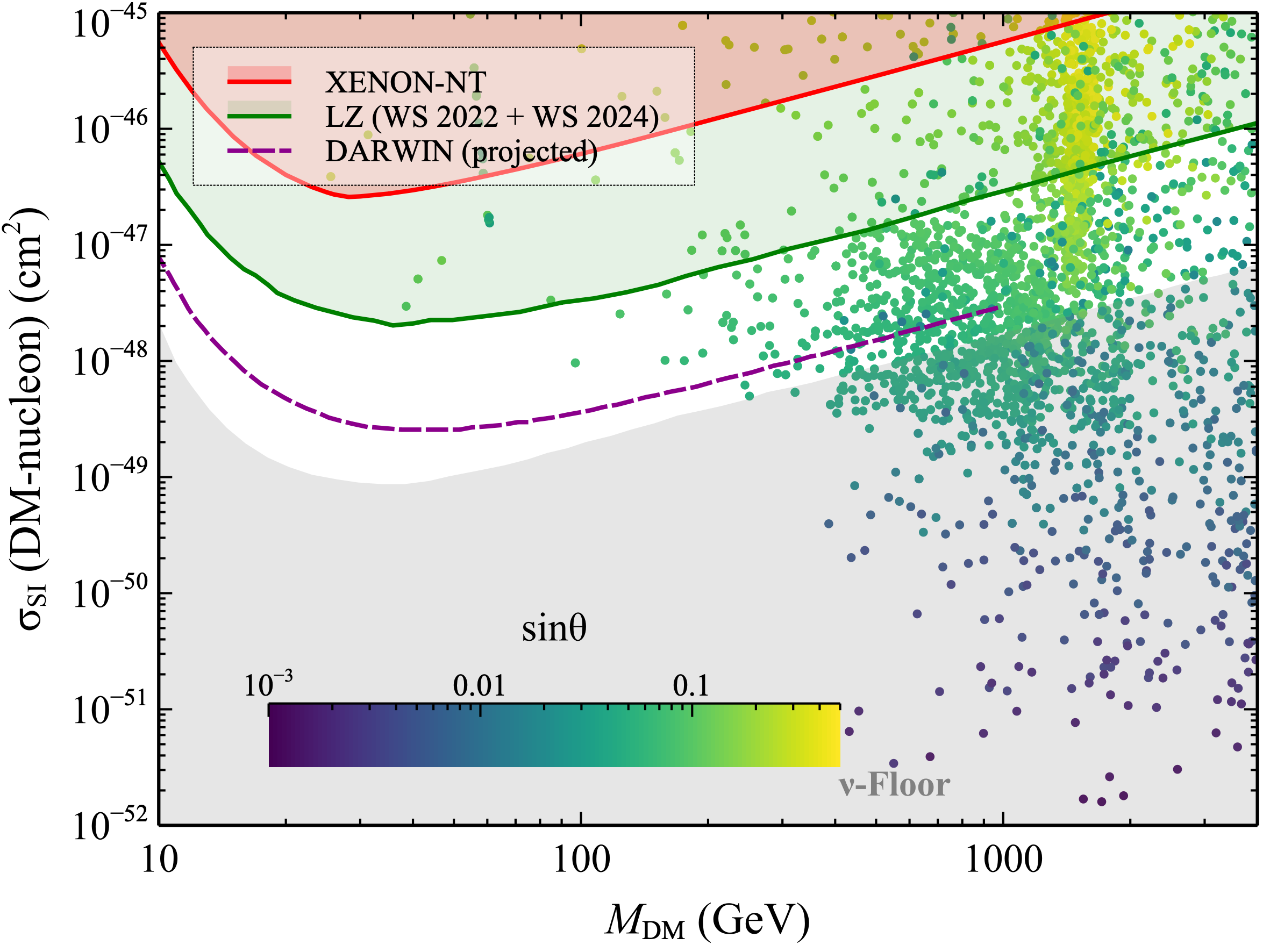}~~
    \includegraphics[scale=0.35]{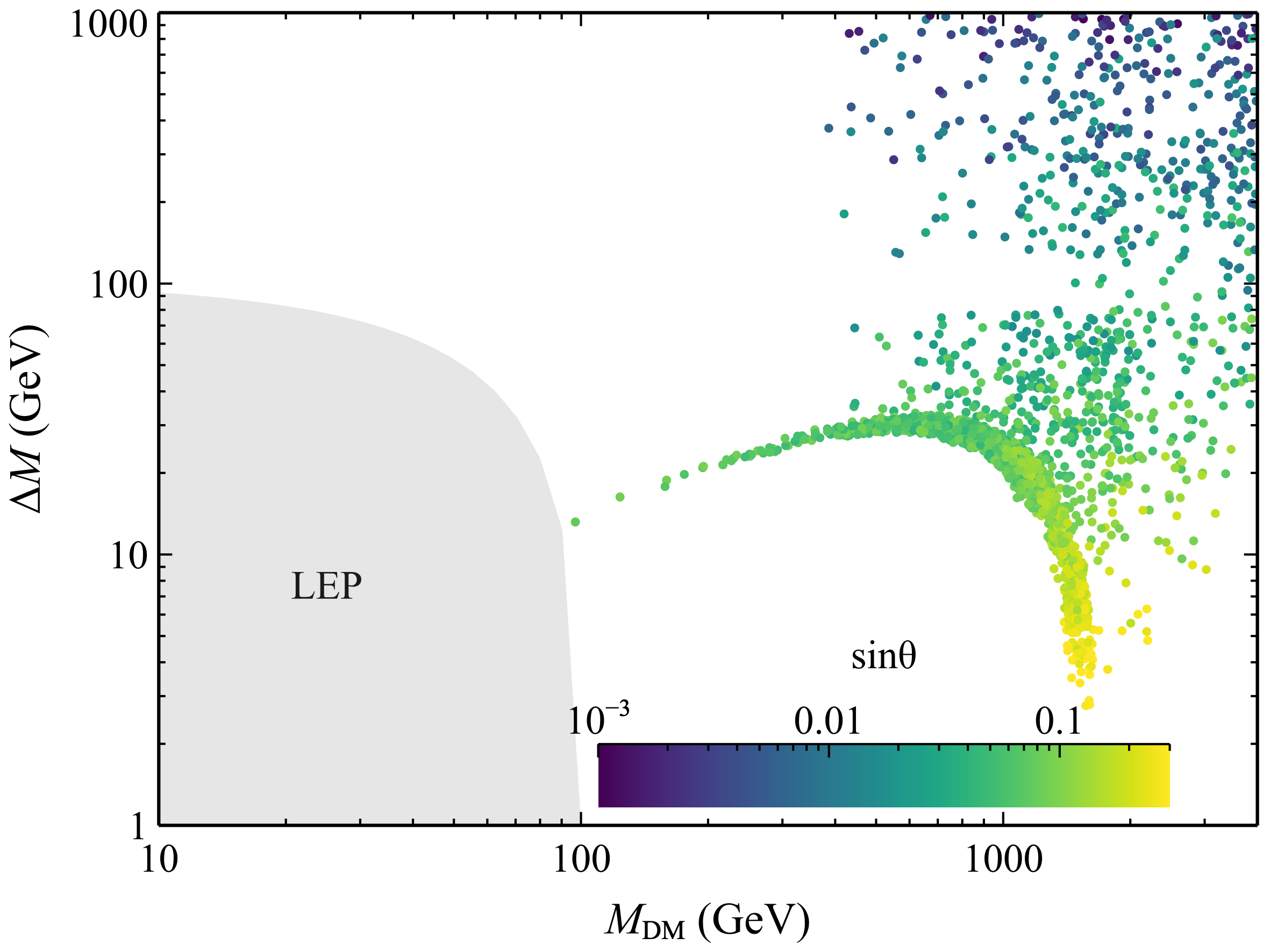}
    \caption{[\textit{Left}]: spin-independent direct detection cross-section as a function of DM mass. [\textit{Right}]: points satisfying correct relic and direct detection from the LZ experiment are shown in the $\Delta{M}-M_{\rm DM}$ plane. The color code represents $\sin\theta$. We note that all these points satisfy the constraints from neutrino mass, $(g-2)_\mu$, and cLFV.}
    \label{fig:dd1}
\end{figure}

In Fig. \ref{fig:dd1} (\textit{left} panel), we show the direct detection cross-section of the relic, constraints from neutrino mass, $(g-2)_\mu$, and cLFV satisfying points as a function of DM mass. The SD mixing angle is shown in color code. We show the current observational constraint from XENONnT \cite{XENON:2023cxc} and LZ \cite{LZ:2022lsv}, along with the projected bound from DARWIN \cite{DARWIN:2016hyl}. The gray colored shaded region represents the $\nu$-floor. In the \textit{right} panel, we show all the points satisfying relic as well as direct detection constraints from the LZ experiment in the plane of $M_{\rm DM}$ and $\Delta M$. In contrast to the pure SDDM model \cite{Paul:2025spm}, this scalar extended setup allows a large viable parameter space even after direct detection constraints. This can be attributed to the new points coming through the DM-$\phi_1$ co-annihilation channel. We find that $M_{\rm DM}\lesssim1000$ GeV, and $\sin\theta\gtrsim0.3$ is disfavored.

%%%%%%%%%%%%%%%%%%%%%%%%%%%%%%%%%%%%%%%%%%%%%%%%%%%%%%%%%%%%%%%%%%%%%%%%%%%%%%%%%%%%%%%%%%
\section{Phase Transition and Gravitational waves}
\label{sec:gw}
%%%%%%%%%%%%%%%%%%%%%%%%%%%%%%%%%%%%%%%%%%%%%%%%%%%%%%%%%%%%%%%%%%%%%%%%%%%%%%%%%%%%%%%%%%
In this model, $\phi_i(i=1,2,3)$  are odd under the imposed $\mathcal{Z}_2$ symmetry, which restricts them from obtaining VEVs. However, their interaction with the SM Higgs can make EWPT to be first order. This section outlines the theoretical framework to realize the first-order phase transition (FOPT) and explores the parameter space of our model that gives rise to strong EWFOPT. Also, we study the GW signals that can be within the reach of current and upcoming space-based GW observatories such as LISA \cite{Baker:2019nia,LISA:2017pwj}, BBO \cite{Adelberger:2005bt,Crowder_2005,Harry_2006}, DECIGO \cite{Yunes:2008tw,Yagi_2011,Kawamura_2006}, AEDGE \cite{AEDGE:2019nxb}, $\mu$Ares \cite{Sesana:2019vho}, etc.

\subsection{Phase transition dynamics}

In order to study the dynamics of EWPT at high temperature, the tree-level SM Higgs potential ($V_0$) alone is not sufficient to realize FOPT. To achieve this, we incorporate the $1$-loop corrected Coleman-Weinberg  potential ($V_{\text{CW}}$) and thermal correction ($V_{\text{th}}$) to $V_0$. To make this perturbatively reliable at high temperatures, we include the daisy resummed potential ($V_{\text{daisy}}$). Since 1-loop corrected potentials introduce the divergence, we need to include counter-terms ($V_{\text{ct}}$) to preserve renormalizability and to ensure that there are no changes in minima in the zero-temperature potential.
Thus, the corresponding effective potential can be given as,
\begin{equation}
    \label{veff}
    V_{\text{eff}}(h,T) = V_0(h) + V_{\text{CW}}(h) +V_{\text{th}}(h,T)+V_{\text{daisy}}(h,T)+V_{\text{ct}}(h).
\end{equation}
In our case, since only the Higgs gets VEV, the tree-level potential in Eq.~(\ref{v0}) modifies to,
\begin{equation}
\label{V0}
    V_0(h)= -\frac{\mu_h^2}{2} h^2 + \frac{\lambda_h}{4} h^4,
\end{equation}
in terms of the background Higgs field ($h$). The Coleman-Weinberg potential, in the Landau gauge and $\overline{MS}$  renormalization scheme, is given by \cite{PhysRevD.7.1888, Quiros:1999jp, PhysRevD.9.1686},
\begin{equation}
    V_{\text{CW}}(h)= \frac{1}{64 \pi^2}\sum_{i} (-1)^{2s_i} n_i M_i^4(h) \left( \log \frac{M_i^2(h)}{Q^2} -C_i\right),
\end{equation}
where $i$ runs for bosons (including Goldstones) and fermions in our model, $s_i$, $n_i$, $M_i(h)$ are spin, number of DoF, and field-dependent mass of the $i$-th particle, respectively. Here, $C_i$'s are 3/2 for scalars and fermions and 5/6 for gauge bosons. The quantity $Q$ represents the renormalization scale of the theory, which we fix to be the mass of the heaviest particle $M_{\phi_3}$. The thermally corrected potential contributing to the effective potential can be written as \cite{PhysRevD.9.3320, Quiros:1999jp},
\begin{equation}
\label{vth}
    V_{\text{th}}(h,T)=\frac{T^4}{2\pi^2}\sum_i (-1)^{2s_i} n_i J_{B/F}\left(\frac{M_i^2(h)}{T^2}\right),
\end{equation}
where $J_{B/F}$ are called thermal functions for bosons/fermions, defined as,
\begin{equation}
    J_{B/F}(x)=\int_0^\infty dy \hspace{1mm} y^2 \log \left(1 \mp e^{-\sqrt{x^2+y^2}} \right).
\end{equation}
To regulate the infrared divergence arising from nearly massless bosons or from high temperatures, we need one-loop thermal correction, called daisy resummation \cite{Arnold:1992rz,Parwani:1991gq}, which is defined as, 
\begin{equation}
    V_{\text{daisy}}(h,T)=-\frac{T}{12\pi}\sum_{i\hspace{0.5mm}\in B} n_i \left[ (M_i^2(h)+\Pi_i(T))^{3/2}-(M_i^2(h))^{3/2}\right],
\end{equation}
where $\Pi_i(T)$ is the thermal (Debye) mass correction for the $i$th bosonic mode. All the field-dependent and thermal masses are shown in Appendix~\ref{subsec:B}.

\subsection{Relevant parameters for GWs}
In the early universe, phase transitions (PTs) proceed via thermal tunneling \cite{Quiros:1999jp,Hindmarsh:2020hop}. The PT tunneling rate is  quantified as $\Gamma(T)\approx T^4 \left(S_3(T)/2 \pi T\right)^{3/2} \exp({-S_3(T)/T})$, where $S_3(T)$ is the $3D$ Euclidean action. Here we will discuss important parameters that we have used to calculate GWs spectrum with the help of publicly available software  \texttt{CosmoTransitions}\footnote{\color{blue}https://github.com/clwainwright/CosmoTransitions }\cite{Wainwright:2011kj}.

When the Universe cools down to the electroweak scale, the effective potential ($V_{\text{eff}}$), which at high temperatures has its minimum at the origin (the unbroken phase), begins to develop a new minimum at a non-zero field value (the broken phase) through a FOPT. The temperature at which the two minima become degenerate is called the critical temperature ($T_c$). Just below this temperature, bubbles of the true vacuum start to form, and they expand and eventually collide, generating GWs. This temperature is called the transition temperature ($T_*$). In the literature, there is no precise definition of the transition temperature, as it varies from case to case \cite{Athron:2023xlk}. The choice of $T_\star$ may change the peak GW amplitude and frequency by several orders of magnitude \cite{Athron:2023rfq}.  For a typical transition with negligible supercooling and reheating, the nucleation temperature $T_n$ is the temperature at which at least one bubble of the true vacuum per Hubble volume starts to nucleate, we will use the most common choice of $T_\star$, i.e, $T_\star \approx T_n $ because it is expected to indicate the start of the phase transition.
One can define the inverse duration of the PT w.r.t Hubble expansion rate, $H_n$ at the nucleation temperature, calculated as \cite{Eichhorn:2020upj},
 \begin{equation}
     \frac{\beta}{H_n} \equiv T_n \frac{d}{d T}\left(\frac{S_3}{T}\right)\Bigg|_{T_n}.
 \end{equation}
 
We can define the strength of the PT as the ratio of the vacuum energy density to that of the radiation bath \cite{Eichhorn:2020upj},
\begin{equation}
    \alpha \equiv \frac{\rho_{\text{vac}}}{\rho_R}\Big |_{T_n} = \frac{1}{\rho_R(T_n)}\left( \Delta V_{\text{eff}}(h,T)-\frac{T}{4}\frac{\partial \Delta V_{\text{eff}}(h,T)}{\partial T}\right)\Bigg |_{T_n},
\end{equation}
where $\rho_R=(g_* \pi^2/30) ~ T^4_n$, and $g_*$ is the relativistic DoF in the radiation bath at $T_n$ and $\Delta V_{\text{eff}}$ is the potential difference between the unbroken and broken phase. 

\begin{figure}[t]
\centering
\includegraphics[scale=.36]{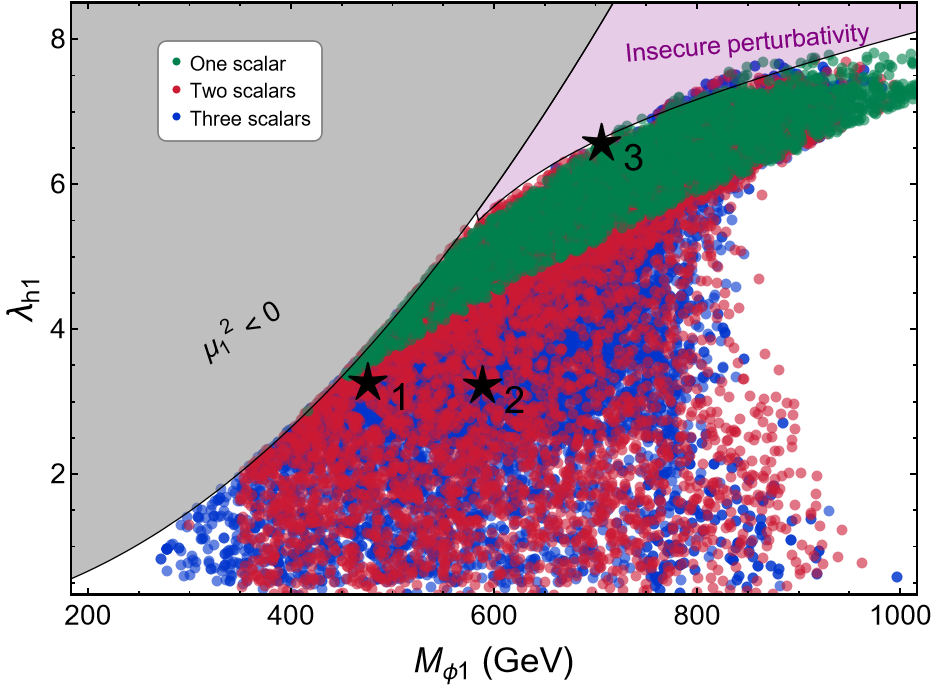}~~
\includegraphics[scale=.37]{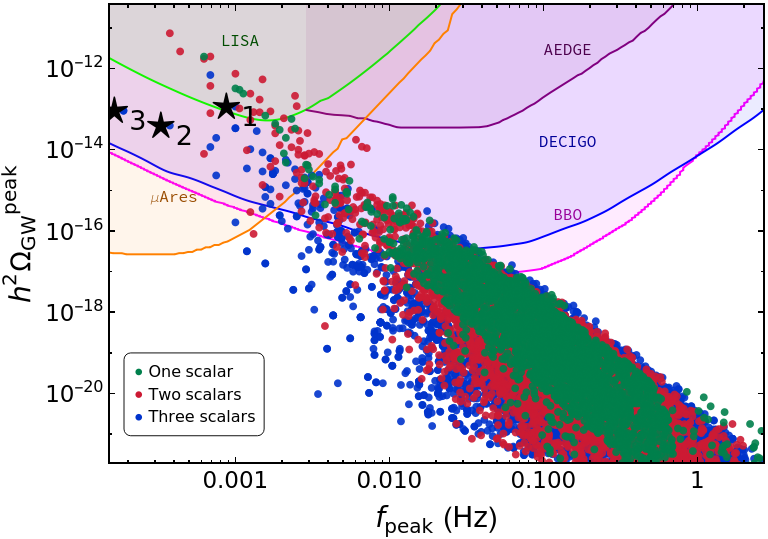}
\caption{[\textit{Left:}] the viable parameter space which gives strong FOPT is shown in the plane of $\lambda_{h1}$ and $M_{\phi_1}$. The forbidden gray region is allowed to give VEV to $\phi$'s, which we don't want here. Each color code is for the addition of each generation of scalars. [\textit{Right:}] the GWs peak amplitude, $h^2 \Omega_{\text{GW}}^{\text{peak}}$ w.r.t peak frequency, $f_{\text{peak}}$, is shown for the strong FOPT, along with the sensitivity curves of various upcoming GW detectors. }
    \label{fig:peakampvspeakfreq}
\end{figure}

The other two important parameters that quantify the fraction of vacuum energy that gets converted into the kinetic energy of the background field $(\rho_{\phi})$ and energy for the bulk motion inside the bubble $(\rho_v)$, respectively \cite{Caprini:2018mtu},
\begin{equation}
    \kappa_{\text{coll}}=\frac{\rho_\phi}{\rho_{\text{vac}}},~~    \kappa_{\text{sw}}=\frac{\rho_v}{\rho_{\text{vac}}},
\end{equation}
and  $\kappa_{\text{turb}} =\epsilon \kappa_{\text{sw}} $ represents the fraction of bulk kinetic energy associated with vertical motions, as opposed to compressional modes.
These can also be expressed in terms of $\alpha$. 
And finally, $v_w$, the velocity of the bubble wall in the rest frame of the fluid far away from any disturbances or collisions, can be calculated using details of hydrodynamics. For our purposes, we assume the runaway bubble, that is, $v_w \approx 1$ for simplicity.
We now discuss the relevant parameter space shown in Fig.~\ref{fig:peakampvspeakfreq}. In the \textit{left} panel, we have shown the available parameter space for strong FOPT in the plane of quartic coupling of Higgs and 1st generation singlet ($\lambda_{h1}$) vs the mass of the singlet scalar ($M_{\phi1}$). During this scanning, we fixed relatively less sensitive parameters, those are $M_\chi=200 ~\text{GeV}, M_\Psi =220~\text{GeV}, \lambda_i\text{'s}=\lambda_{12}=\lambda_{23}=\lambda_{13}=10^{-4},\sin\theta=10^{-3}$, while the ranges of other parameters that take part in random scanning are $M_{\phi_1}=[200-1000]~\text{GeV}, M_{\phi_2}=[200-2000]~\text{GeV},M_{\phi_3}=[200-3000]~\text{GeV}$ and for couplings, $\lambda_{hi}\text{'s}=[0.1 - 10]$.

The gray region allows $\phi_i$'s to get VEV, but in our model, we restrict it by imposing  $\mathcal{Z}_2$  odd symmetry on $\phi_i$'s. It is known from previous studies \cite{Mirzaie:2025bzn,Braconi:2018gxo,PhysRevD.99.015035}, that only the singlet extension of the SM can provide the necessary recipe for FOPT, but may not be strong if it is extended by fermions, which contribute to increasing quadratic couplings in the effective potential, favoring symmetry restoration. Therefore, we add additional scalars to enhance our parameter space for strong FOPT $( v_c/T_c\gtrsim 1$ \cite{Quiros:1999jp,Oikonomou:2024jms}). As we can see from \textit{left} panel, the viable space for a single scalar is very small compared to the two other scenarios, which may not provide enough space to probe our model by GW signals. Notice that the addition of the second and third scalars provides an almost identical parameter space. However, we add the third scalar for the sake of completeness of the three neutrino mass eigenstate generation via a radiative loop. The purple shaded region indicates the perturbative bound, calculated from the zero-temperature one-loop potential. We have estimated the 1-loop correction to the Higgs quartic coupling ($\Delta \lambda_h$) in the asymptotic limit, and it should not exceed unity. This is an approximate boundary, as it depends on other model parameters. For the three scalar case, we have calculated $\Delta \lambda_h$ for each point in the scatter plot and found $(\Delta \lambda_{h})_{\rm max} \approx 0.23 $, as shown in the plot.   It is noticed that for $\lambda_{h1}\lesssim 7$, our results can be trusted to provide the parameter space for strong EWFOPT.

In the \textit{right} panel of Fig. \ref{fig:peakampvspeakfreq}, we have shown the peak amplitude of GWs, $h^2\Omega_{\rm GW}^{\rm peak}$ versus peak frequency, $f_{\rm peak}$, along with the projected sensitivity curves of upcoming GW observatories. All relevant formulas are provided in Appendix \ref{app:FOPT}. The color codes are mentioned in the plot itself. It is noticed that, across all scenarios, the experimental reach is essentially unchanged, although the scatter points are somewhat more widely distributed than in the single-scalar case, potentially improving detection prospects. In the frequency range 0.1 mHz to 10 mHz, our model can be well-probed by BBO, DECIGO, $\mu$Ares, and LISA. However, for AEDGE, the amplitude is too weak to reach the high-frequency regime.

%%%%%%%%%%%%%%%%%%%%%%%%%%%%%%%%%%%%%%%%%%%%%%%%%%%%%%%%%%%%%%%%%%%%%%%%%%%%%%%%%%%%%%%%%%    
\section{Results and discussion}\label{sec:resultand discussion}

In the above sections (\ref{sbsec:numass}, \ref{sbsec:muong-2}, \ref{sbsec:lfv}, \ref{sec:dmpheno}), we discussed the DM parameter space extensively while satisfying the $\nu$-mass followed by $(g-2)_\mu$ and cLFV interaction. We also have provided an analysis of the generation of GW due to the modification of the Higgs scalar potential in the presence of $\phi_{i}(i=1,2,3)$s in section \ref{sec:gw}. In this section, we provide a common parameter space that satisfies all the above-discussed phenomena.

\begin{figure}[h]
    \centering
    \includegraphics[scale=0.5]{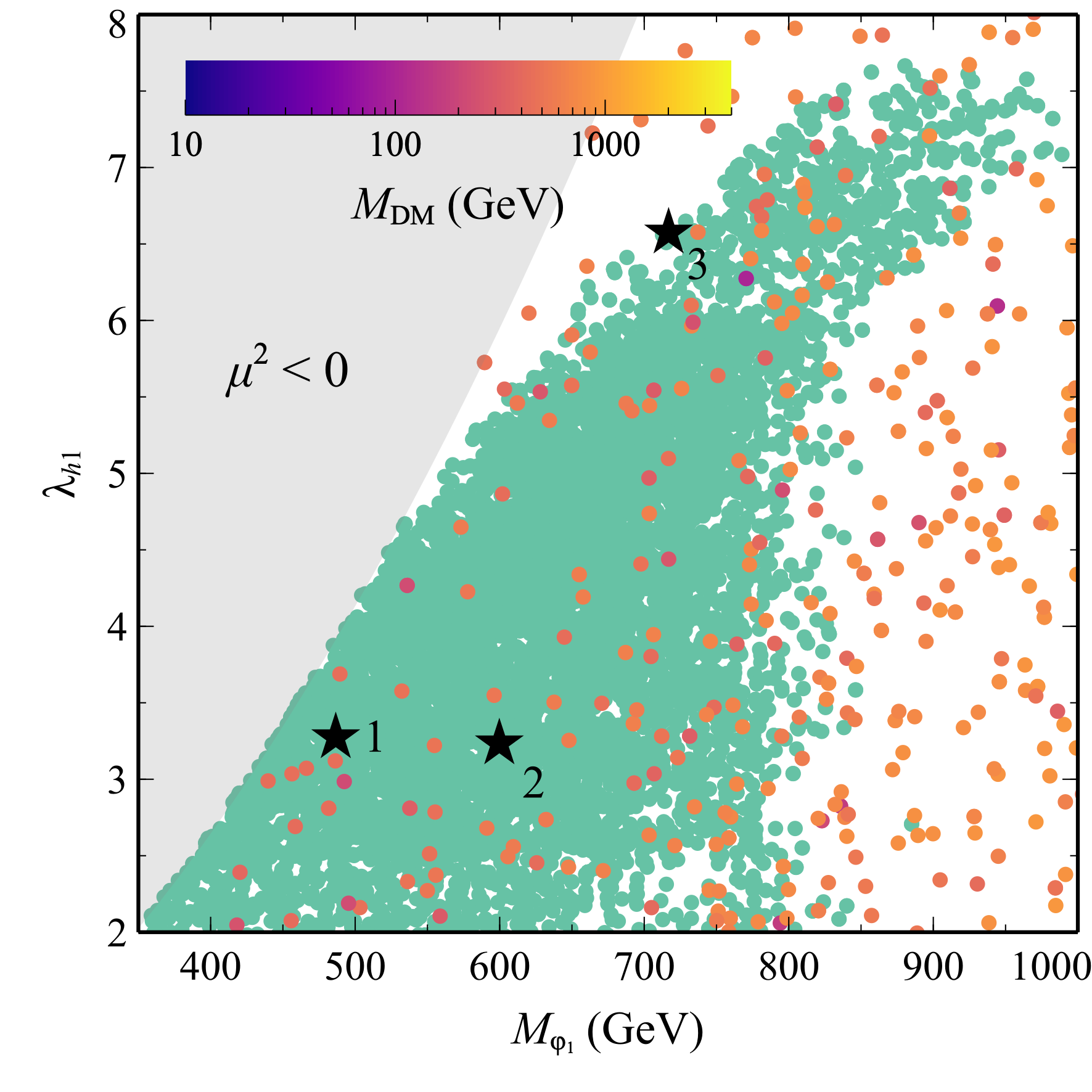}~~
\includegraphics[scale=0.5]{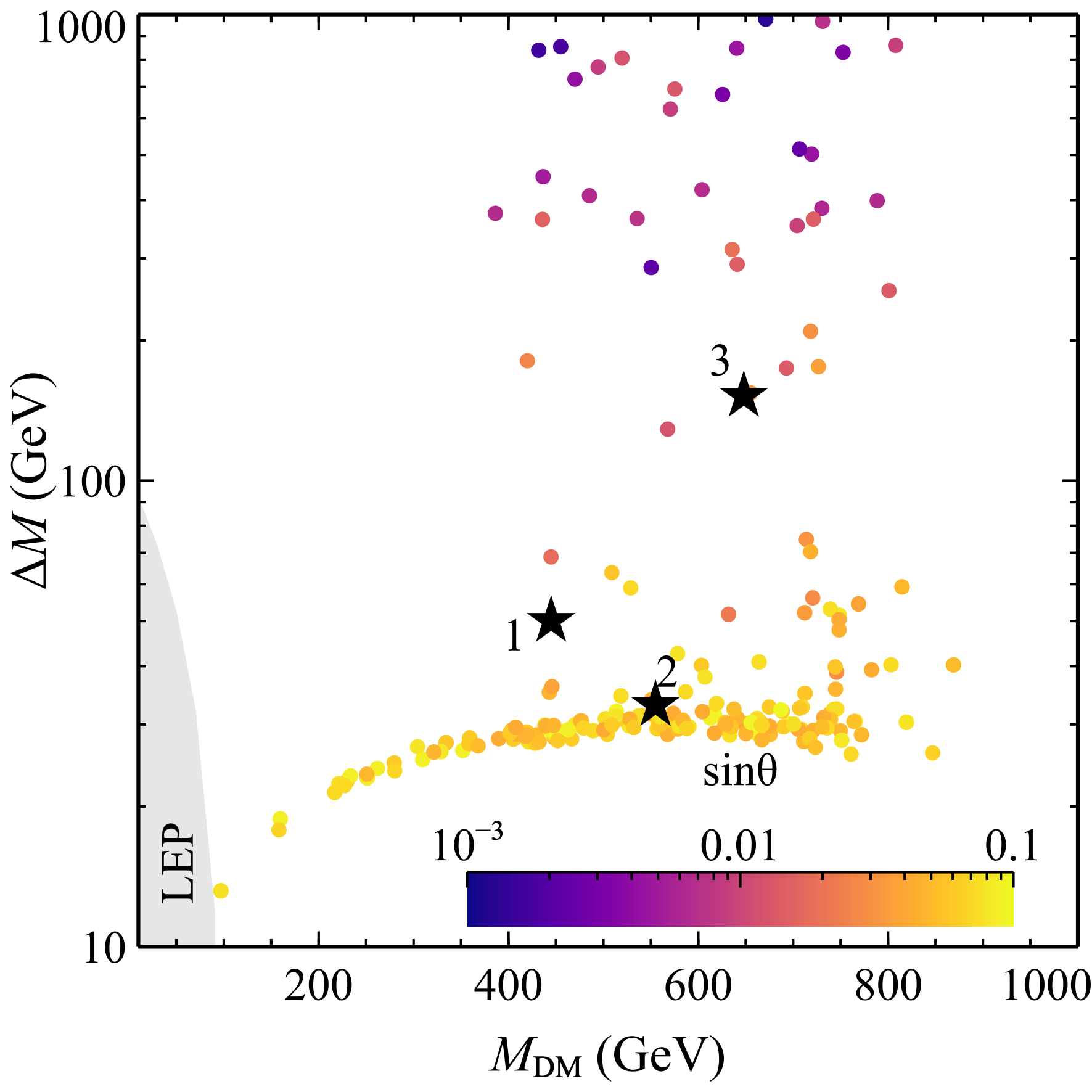}
    \caption{[\textit{Left:}] gradient colored relic satisfying points are shown in the plane of $M_{\phi_1}$ and $\lambda_{h1}$. The blue  colored points represent the FOPT satisfying points considering three scalars. [\textit{Right:}] correct relic satisfying points overlapping with the FOPT parameter space are shown in the $\Delta{M}-M_{\rm DM}$ plane. The $\sin\theta$ is shown in the color code. The gray shaded region represents the LEP bound.}
    \label{fig:summary1}
\end{figure}

In the \textit{left} panel of Fig. \ref{fig:summary1}, we showcase the parameter space that gives rise to GW in the plane of $M_{\phi_1}$ and $\lambda_{h1}$. We also show the points satisfying neutrino mass, muon $(g-2)$, cLFV, DM relic, and direct detection constraints in the same plane with $M_{\rm DM}$ in color code. In the \textit{right} panel of Fig. \ref{fig:summary1}, we show the points satisfying both the FOPT and the other phenomenological constraints discussed above are shown in the plane of $\Delta{M}-M_{\rm DM}$ with $\sin\theta$ in the color code. We see that $\sin\theta$ up to 0.1 is allowed by the combined phenomenological constraints.

In table \ref{tab:tab3}, we show three benchmark points that satisfy neutrino mass, muon $g-2$, cLFV, DM relic, and direct detection simultaneously and give observable gravitational wave signatures through the first-order electroweak phase transition. These two points are shown with black ``$\bigstar$'' in the Fig. \ref{fig:peakampvspeakfreq}. The same BPs are also shown in Fig. \ref{fig:summary1}.

\begin{table}[h!]
%\centering
\begin{minipage}{0.40\textwidth}
\small
\centering
\renewcommand{\arraystretch}{1.3}
\resizebox{6cm}{!}{
\begin{tabular}{|c|c|c|c|}
\hline
\textbf{Parameters} & \textbf{BP1}&\textbf{BP2} &\textbf{BP3}\\
\hline
$M_{\rm DM}$ (GeV) & 445 & 555 & 648  \\
\hline
$\Delta M$ (GeV) & 50 & 33 & 152 \\
\hline
$\sin\theta$ & 0.1 & 0.05  & 0.035 \\
\hline
$M_{\phi_1}$ (GeV) & 468.56 & 599.75 & 716.86  \\
\hline
$M_{\phi_2}$ (GeV) & 511.39 & 625.32 &863.09  \\
\hline
$M_{\phi_3}$ (GeV) & 536.31 & 650.89 & 932.67  \\
\hline
$\lambda_{h1}$ & 3.274 & 3.234 & 6.574  \\
\hline
$\lambda_{h2}$ & 4.021 & 3.095 & 1.643  \\
\hline
$\lambda_{h3}$ & 3.568 & 5.73 & 3.661  \\
\hline
\end{tabular}}
%\caption{Benchmark Point 1}
\end{minipage}
\hspace{0.02\textwidth}
\begin{minipage}{0.60\textwidth}
\small
%\centering
\renewcommand{\arraystretch}{1.34}
\resizebox{8cm}{!}{
\begin{tabular}{|c|c|c|c|}
\hline 
\textbf{\shortstack{Constraints$/$ \\  Observables }} & \textbf{BP1}&\textbf{BP2}&\textbf{BP3} \\
\hline
$\nu$ {\rm mass} &  \textcolor{blue}{\CheckmarkBold} &\textcolor{blue}{\CheckmarkBold}&  \textcolor{blue}{\CheckmarkBold} \\
\hline
$(g-2)_\mu$ & \textcolor{blue}{\CheckmarkBold} & \textcolor{blue}{\CheckmarkBold}&  \textcolor{blue}{\CheckmarkBold}   \\
\hline
{\rm cLFV} & \textcolor{blue}{\CheckmarkBold} & \textcolor{blue}{\CheckmarkBold}&  \textcolor{blue}{\CheckmarkBold}  \\
\hline
{\rm DM relic} & \textcolor{blue}{\CheckmarkBold} & \textcolor{blue}{\CheckmarkBold}&  \textcolor{blue}{\CheckmarkBold}   \\
\hline
{\rm DD} & \textcolor{blue}{\CheckmarkBold} & \textcolor{blue}{\CheckmarkBold} &  \textcolor{blue}{\CheckmarkBold}  \\
\hline
$\alpha$ & 0.05499 & 0.02352 &0.02274  \\
\hline
$\beta/H_n$ & 66.35 & 19.23 & 9.63  \\
\hline
$h^2\Omega_{\rm GW}^{\rm peak}$ & $1.15\times10^{-13}$ & $3.39\times10^{-14}$ & $9.11\times10^{-14}$   \\
\hline
$f_{\rm peak}$ (Hz)& $10^{-3}$ & $3.75 \times 10^{-4}$& $1.87\times10^{-4}$ \\
\hline
\end{tabular}}
\end{minipage}
\caption{Benchmark points satisfying neutrino mass, muon $g-2$, cLFV, DM relic, direct detection, and giving observable gravitational wave signatures through first-order electroweak phase transition. }
		\label{tab:tab3}
\end{table}

As already discussed in the above sections, the neutrino mass, muon ($g-2$), and cLFV combinedly give a lower bound on the $\sin\theta$ of $\mathcal{O}(10^{-3})$. The successful EWFOPT requires $M_{\phi_1}\lesssim1$ TeV. Now this gives an upper bound on the DM mass to be $M_{\rm DM}\lesssim1$ TeV. This results in an upper limit on the $\sin\theta$ of approximately $0.1$, which can be easily seen from Fig. \ref{fig:dd1}. 
\begin{figure}[h]
    \centering   \includegraphics[scale=0.5]{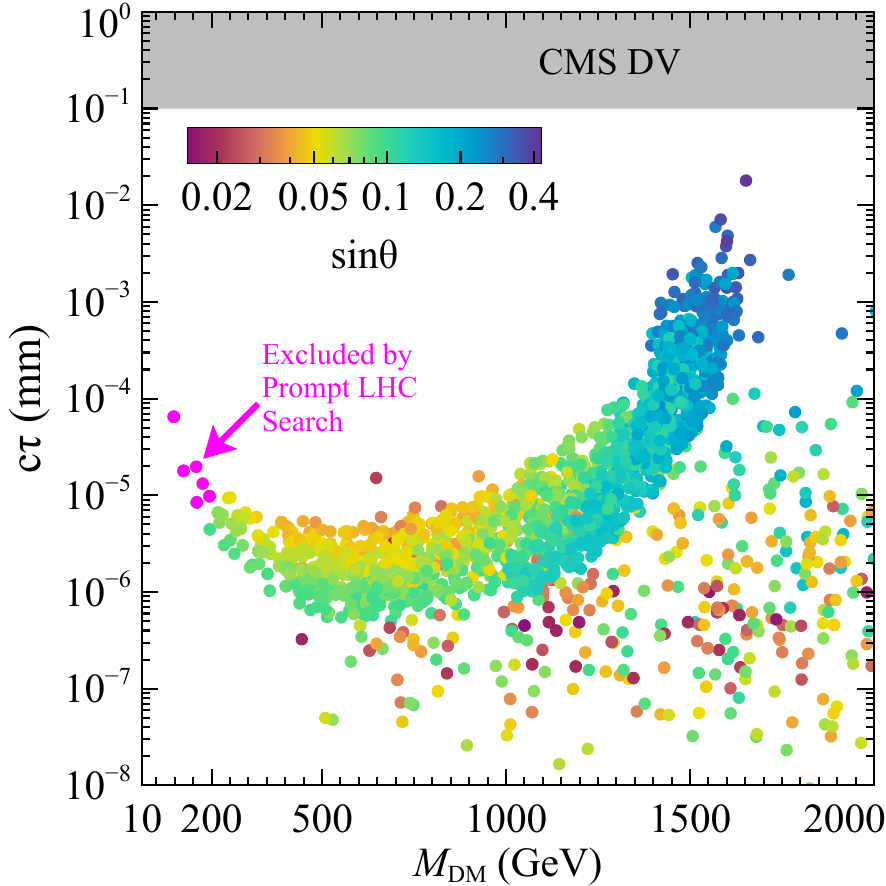}
    \caption{Decay length of the doublet fermion as a function of DM mass with $\sin\theta$ in the color code. The magenta colored points are excluded by the prompt LHC searches \cite{CMS:2021cox,ATLAS:2021moa}.}
    \label{fig:ctau}
\end{figure}

It is worth noting that doublet fermions can be produced at the collider due to their gauge interactions. Once they are produced, the charged component of the doublet fermion, $\psi^\pm$ can decay to DM and charged lepton and neutrino if the mass splitting is $\Delta{M}\lesssim80.4$ GeV via an off shell W boson. In Fig. \ref{fig:ctau}, we show the decay length of $\psi^\pm$ as a function of DM mass. The color code represents $\sin\theta$. These points correspond to the same points as in the \textit{right} panel of Fig. \ref{fig:dd1}. The CMS is sensitive to a decay length in the range of 0.1 mm to 1000 mm, which is shown with the gray shaded region \cite{CMS:2024trg}. We note that the scatter points are below the sensitivity range of CMS, which is because the $\sin\theta$ allowed here are $\gtrsim10^{-2}$ in contrast to the pure SDDM model, where $\sin\theta$ was allowed to vary in the range of $10^{-7}\leq\sin\theta\leq0.16$ \cite{Paul:2025spm}. If the decay length of the charged doublet is shorter than ($10^{-4}$ m), it decays before entering the detector volume, resulting in a prompt decay signature at the LHC. Limits from CMS \cite{CMS:2021cox} and ATLAS \cite{ATLAS:2021moa} using $pp\rightarrow 3l+{E^{\rm miss}_T}$ (${E^{\rm miss}_T}$ denotes the missing transverse energy) exclude the magenta colored points in Fig. \ref{fig:ctau}.

%%%%%%%%%%%%%%%%%%%%%%%%%%%%%%%%%%%%%%%%%%%%%%%%%%%%%%%%

%%%%%%%%%%%%%%%%%%%%%%%%%%%%%%%%%%%%%%%%%%%%%%%%%%%%%%%%%%%%%%%%%%%%%%%%%%%%%%%%%%%%%%%%%%    
\section{Conclusion}\label{sec:concl}
In this work, we have explored a radiative neutrino mass model where the interplay among $\mathcal{Z}_2$-odd singlet and doublet fermions, together with three generations of scalar singlets, gives rise to a unified framework connecting neutrino mass, dark matter, and gravitational wave phenomenology. The singlet–doublet fermion mixing naturally leads to a viable Majorana dark matter candidate, while the scalar singlets significantly alter the Higgs potential, enabling a strong first-order electroweak phase transition that can generate detectable gravitational wave signals. Furthermore, the same new states contribute to neutrino mass generation at one loop and can induce observable effects in $(g-2)_\mu$ and charged lepton flavor violation processes. The direct detection experiment excludes the singlet-doublet mixing angle $\sin\theta>0.3$. The FOPT together with constraints from neutrino mass, muon ($g-2$), cLFV, DM relic, and direct detection favors a DM mass in the range of $100{~\rm GeV}\lesssim M_{\rm DM}\lesssim 900$ GeV and a SD mixing angle in the range of $10^{-3}\lesssim\sin\theta\lesssim0.1$.

%%%%%%%%%%%%%%%%%%%%%%%%%%%%%%%%%%%%%%%%%%%%%%%%%%%%%%%%%%%%%%%%%%%%%%%%%%%%%%%%%%%%%%%%%%
\section*{Acknowledgments}
U.K.D. acknowledges support from the Anusandhan National Research Foundation (ANRF), Government of India under Grant Reference No.~CRG/2023/003769. P.K.P. would like to acknowledge the Ministry of Education, Government of India, for providing financial support for his research via the Prime Minister’s Research Fellowship (PMRF) scheme. SKM also wishes to thank Arka Bhattacharyya, Kaustav Mukherjee, and Indra Kumar Banerjee for their guidance in implementing \texttt{CosmoTransitions}. We thank the hospitality of the organizers of Phoenix-2025 at IIT Hyderabad where the project was initiated.

%%%%%%%%%%%%%%%%%%%%%%%%%%%%%%%%%%%%%%%%%

%%%%%%%%%%%%%%%%%%%%%%%%%%%%%%%%%%%%%%%%%

%
\appendix
\section{Radiative neutrino mass}\label{app:numass_matrix}
The matrix element for the radiative neutrino mass diagram presented in Fig.~\ref{fig:numass} is given by
\begin{eqnarray}
    \overline{(\nu_L)^C}_\alpha(\mathcal{M}_\nu)_{\alpha\beta}{\nu_L}_\beta&=&\overline{(\nu)^C}_\alpha\left[\int \frac{d^4q}{(2\pi)^4} ~\frac{v^2}{2}~y_{\alpha }y_{\beta }y_\chi^2P_L\frac{\cancel{q}+M_\Psi}{q^2-M_\Psi^2}P_R\frac{\cancel{q}+M_\chi}{q^2-M_\chi^2}P_R\times\right.\nonumber\\&&~~~~~~~~~~~~~~~~~~~~~~~~~~~~~~~~~~~~\left.\frac{\cancel{q}+M_\Psi}{q^2-M_\Psi^2}P_L\frac{1}{q^2-M_\phi^2}\right]{\nu}_\beta,
\end{eqnarray}
where $q$ is the loop momentum.
The above expression simplifies to:
\begin{align}
    (\mathcal{M}_\nu)_{\alpha\beta}=&\frac{v^2}{32\pi^4}~y_{\alpha }y_{\beta }y_\chi^2\int d^4q ~\frac{4q^2M_\chi}{(q^2-M_\Psi^2)^2(q^2-M_\chi^2)(q^2-M_\phi^2)}\nonumber\\
    &=\frac{v^2}{8\pi^2}~y_{\alpha }y_{\beta }y_\chi^2 M_\chi~\left[\frac{M_\chi^4}{(M_\chi^2-M_\phi^2)(M_\chi^2-M_\Psi^2)^2}\log\left[\frac{M_\chi^2}{M_\Psi^2}\right]\right.\nonumber\\
    &~~~~~~~\left.+\frac{M_\Psi^2}{(M_\chi^2-M_\Psi^2)(M_\phi^2-M_\Psi^2)}-\frac{M_\phi^4}{(M_\chi^2-M_\phi^2)(M_\phi^2-M_\Psi^2)^2}\log\left[\frac{M_\phi^2}{M_\Psi^2}\right]\right]
\end{align}
\section{Feynman Diagrams of the processes involved in relic density}\label{app:feyndiags}

\subsection*{Annihilation processes of DM}
\begin{figure}[H]
    \centering
    \includegraphics[scale=0.9]{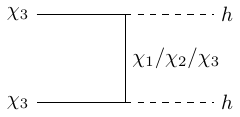}
    \includegraphics[scale=0.9]{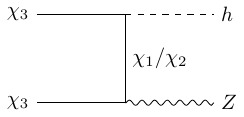}
    \includegraphics[scale=0.9]{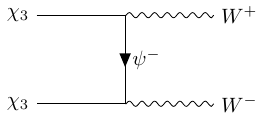}
    \includegraphics[scale=0.9]{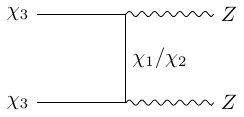}
    \includegraphics[scale=0.7]{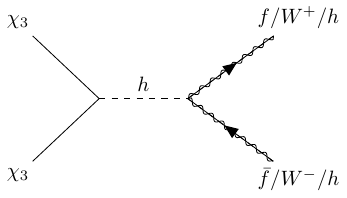}
    \caption{DM annihilating to the SM particles through 1100 processes.}
    \label{fig:annDM}
\end{figure}

Here, we have provided the Feynman diagram of annihilation (corresponds to 1100 processes) of sector 1 particle (i.e. DM or $\chi_3$) to sector 0 particles (i.e SM particles) in Fig. \ref{fig:annDM}

\subsection*{Co-annihilation between DM and doublet fermions and singlet scalar}
\begin{figure}[H]
    \centering
    \includegraphics[scale=0.7]{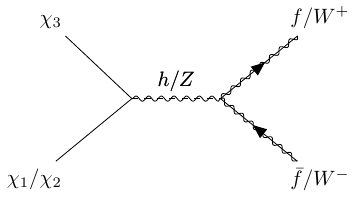}
    \includegraphics[scale=0.7]{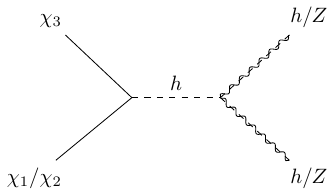}
    \includegraphics[scale=0.7]{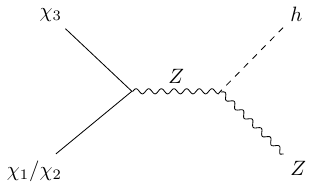}
    \includegraphics[scale=0.9]{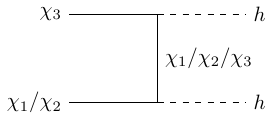}
    \includegraphics[scale=0.9]{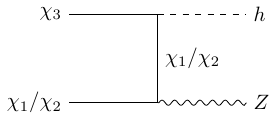}
    \includegraphics[scale=0.9]{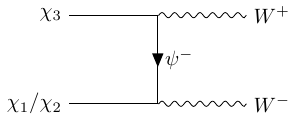}
    \includegraphics[scale=0.9]{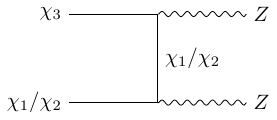}
    \includegraphics[scale=0.9]{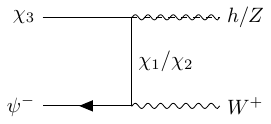}
    \includegraphics[scale=0.9]{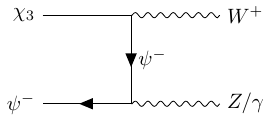}
    \includegraphics[scale=0.7]{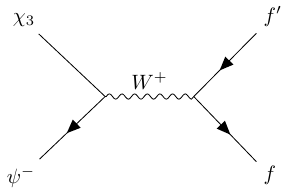}
    \includegraphics[scale=0.7]{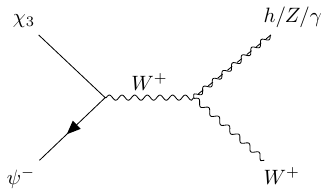}
    \includegraphics[scale=0.9]{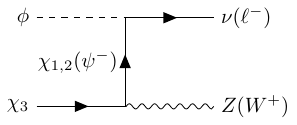}
    \caption{DM co-annihilating with the doublet components and singlet scalars to the SM particles through 1200 processes.}
    \label{fig:coannDM1}
\end{figure}

The co-annihilation (corresponds to 1200 processes) among the sector 1 and sector 2 particles (i.e. $\chi_1,\chi_2,\psi^{-}$ and $\phi_{1,2,3}$) to SM particles are shown in Fig. \ref{fig:coannDM1}

\subsection*{Annihilation and co-annihilation among the sector 2 particles}
\begin{figure}[H]
    \centering
    \includegraphics[scale=0.9]{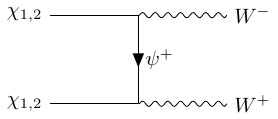}
    \includegraphics[scale=0.9]{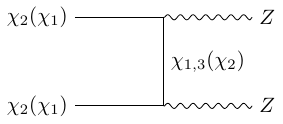}
    \includegraphics[scale=0.7]{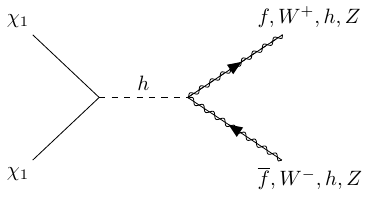}
    \includegraphics[scale=0.9]{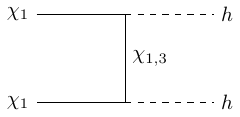}
    \includegraphics[scale=0.9]{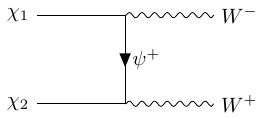}
    \includegraphics[scale=0.7]{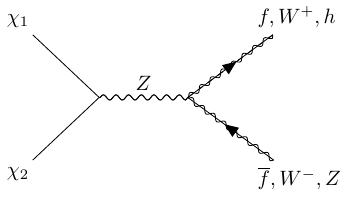}
    \includegraphics[scale=0.9]{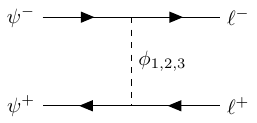}
    \includegraphics[scale=0.9]{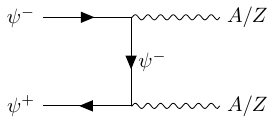}
    \includegraphics[scale=0.7]{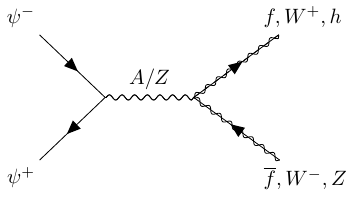}
    \includegraphics[scale=0.9]{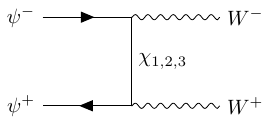}
    \includegraphics[scale=0.9]{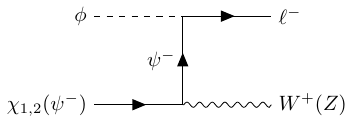}
    \includegraphics[scale=0.7]{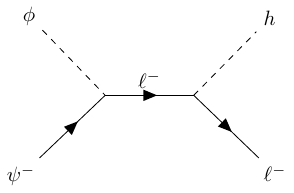}
    \includegraphics[scale=0.7]{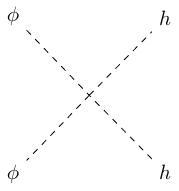}
    \includegraphics[scale=0.9]{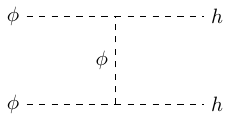}
    \includegraphics[scale=0.9]{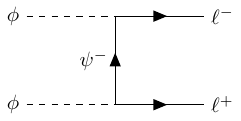}
    \includegraphics[scale=0.7]{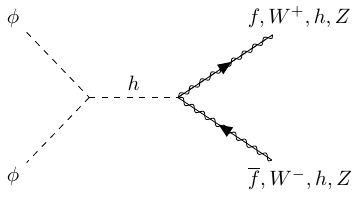}
    \caption{Annihilation and co-annihilation among the sector 2 particles to SM particles corresponding to the 2200 processes.}
    \label{fig:annDM2}
\end{figure}

Annihilation and co-annihilation (corresponding to 2200 processes) among the sector 2 particles to SM particles are given in Fig. \ref{fig:annDM2}.

\section{Details of FOPT and GW formulas}
\label{app:FOPT}
\subsection*{Sources of gravitational waves}
The power spectrum of GWs from FOPT, generated from three main mechanisms: bubble collisions, sound waves, and magnetohydrodynamics turbulence in the plasma, is \cite{Caprini:2015zlo},
\begin{equation}
    h^2 \Omega_{\text{GW}}(f)=h^2 \Omega_{\text{coll}}(f) + h^2 \Omega_{\text{sw}}(f) + h^2 \Omega_{\text{turb}}(f),
\end{equation}
where each term explicitly depends on PT parameters $\alpha, \beta/H_n, T_n, v_w$, which we discussed previously. Moreover, the details of each term are discussed in the following.

\subsubsection*{Collision of bubble walls}
Among the three sources of GWs, this is the weakest in amplitude. Assuming the envelope approximation \cite{PhysRevD.45.4514,PhysRevLett.69.2026} used in numerical simulations, the contribution to the GWs \cite{Caprini:2015zlo} is given by, 
\begin{equation}
    h^2\Omega_{\text{coll}}(f)=1.67\times10^{-5} \left (\frac{H_n}{\beta}\right )^{2} \left( \frac{\alpha ~\kappa_\phi }{1+\alpha}\right)^2 \left(\frac{100}{g_\star(T_n)}\right)^\frac{1}{3} \left(\frac{0.11 \hspace{0.1cm}v_w^3}{0.42 ~ +v_w^2}\right)  \left(\frac{3.8~\left(\frac{f}{f_{\text{coll}}}\right)^{2.8}}{1+2.8~\left(\frac{f}{f_{\text{coll}}}\right)^{3.8}}\right),
\end{equation}
where the peak frequency after redshift is, 
\begin{equation}
f_{\text{coll}} = 1.65\times 10^{-5} \hspace{0.1cm} \text{Hz}\left (\frac{\beta}{H_n}\right )\left(\frac{T_n}{100}\right)\left(\frac{0.62}{1.8 -0.1\hspace{0.1cm}v_w+v_w^2}\right)
\end{equation}
and the fractional vacuum energy converted to bubble collision, $\kappa_\phi=\frac{0.715\hspace{0.1cm}+\frac{4}{27}\hspace{0.05cm}\sqrt{1.5 \hspace{0.1cm}\alpha}}{1+ 0.715 \hspace{0.1cm}\alpha}.$

\subsubsection*{Sound wave}
The pressure difference is created on either side of the bubble to propagate it through the plasma to create bulk motion in the fluid in the form of GWs. For the generic values of $v_w$, the numerical results are well fitted with the following formula \cite{Ellis:2019oqb,Hindmarsh:2013xza}  
\begin{equation}
    h^2\Omega_{\text{sw}}(f) = 2.65\times10^{-6} ~G\left (\frac{H_n}{\beta}\right ) v_w \left( \frac{\alpha ~ \kappa_{\text{sw}} }{1+\alpha}\right)^2 \left(\frac{100}{g_\star(T_n)}\right)^\frac{1}{3}  \left(\frac{7}{4+3~\left(\frac{f}{f_{\text{sw}}}\right)^2}\right)^\frac{7}{2}~\left(\frac{f}{f_{\text{sw}}}\right)^3,
\end{equation}
where the redshifted peak frequency is given by,
\begin{equation}
    f_{\text{sw}} = 1.9\times 10^{-5}\hspace{0.1cm} \text{Hz}\left (\frac{\beta}{H_n}\right )\left(\frac{T_n}{100}\right)\left(\frac{g_\star(T_n)}{100}\right)^\frac{1}{6} \left(\frac{1}{v_w}\right),
\end{equation}
the fractional vacuum energy converted to bubble collision, $\kappa_{\text{sw}}=\frac{\alpha}{\alpha + 0.73 +0.083\sqrt{\alpha}}$,
lastly the factor, $G=1-1/\sqrt{1+2(8\pi)^{1/3}  (H_n/\beta)v_w\sqrt{4(1+\alpha)/(3\alpha \kappa_{sw})} }$, is a suppression factor \cite{Lewicki_2022} that arises due to the finite lifetime of the sound wave.

\subsubsection*{Turbulence}
In the presence of a primordial magnetic field, when the bubble propagates into a fully ionized plasma, it creates turbulent motion and contributes to the GW spectrum. Using magnetohydrodynamics (MHD) modeling, the following amplitude is calculated as \cite{Binetruy:2012ze, Caprini:2015zlo}
 \begin{equation}
h^2\Omega_{\text{turb}}(f) = 3.35\times10^{-4}~\left (\frac{H_n}{\beta}\right )v_w \left( \frac{\alpha ~\kappa_{\text{turb}} }{1+\alpha}\right)^\frac{3}{2} \left(\frac{100}{g_\star(T_n)}\right)^\frac{1}{3}\left(\frac{\left(\frac{f}{f_{\text{turb}}}\right)^3}{1+\left(1+\frac{8\pi f}{h_\star(T_n)}\right)~\left(\frac{f}{f_{\text{turb}}}\right)^\frac{11}{3}}\right)
 \end{equation}
where the peak frequency, 
\begin{equation}   
f_{\text{turb}} = 2.7\times 10^{-5}\hspace{0.1cm} \text{Hz}\left(\frac{T_n}{100}\right)\left(\frac{g_\star(T_n)}{100}\right)^\frac{1}{6} \left(\frac{1}{v_w}\right),
\end{equation}
the fractional vacuum energy converted to bubble collision, $\kappa_{\text{sw}}=\epsilon \kappa_{\text{sw}}$ (we take $\epsilon=0.1$ \cite{Borah_2023}), and the inverse Hubble time at the production of GW, redshifted today is \cite{Caprini:2015zlo},
$h_\star(T_n)=1.65\times10^{-5}\hspace{0.1cm} \text{Hz}\left(\frac{T_n}{100}\right)\left(\frac{g_\star(T_n)}{100}\right)^\frac{1}{6}$.

\subsection*{Field dependent masses}
\label{subsec:B}
 If we express the scalar potential in Eq.(\ref{v0}) in terms of the fields $h, G_0, G_\pm, %\text{(Goldstone bosons in Higgs doublet)}, 
 \phi_{1,2,3}$ and calculate $M^2_{i j}=\frac{\partial^2 V}{\partial x_i \partial x_j}$, where $x_i \in \{h, G_0, G_\pm, \phi_{1,2,3}\}$, we find the following
\begin{align}
     M_h^2(h)=&\lambda_h(3h^2-v^2),\\
     M^2_{\phi_i}(h) =& \lambda_{hi}(h^2 -v^2) + M^2_{\phi_i}, \\ 
     M^2_{G_0}(h) = &M^2_{G_\pm}(h)= \lambda_h (h^2 -v^2),
 \end{align}
 where, $i$ runs from 1 to 3 and $v$ refers to zero temperature Higgs VEV.
The field-dependent fermionic masses are given by
 \begin{equation}
     M_t(h)=\frac{y_t}{\sqrt{2}}h,
 \end{equation}
 as only the top quark mass is relevant here, and recall from Eq.(\ref{fermion masses})
 \begin{eqnarray}
M_{\chi_1} (h)&=&M_\Psi\cos^2\theta+M_\chi\sin^2\theta+\frac{y_{\chi} h }{\sqrt{2}} \sin2\theta,\\
M_{\chi_3}(h)&=&M_\Psi\sin^2\theta+M_\chi\cos^2\theta-\frac{y_{\chi} h }{\sqrt{2}}\sin2\theta.
\end{eqnarray}
Similarly, the gauge boson masses are expressed as
\begin{align}
    M_Z (h) & = \frac{h}{2}\sqrt{g^2 +g^{\prime 2}}\\
    M_W(h) & = \frac{g h}{2},
\end{align}
 where $g, g^\prime$ has the usual meaning in the literature.

\subsection*{Counter terms}
\label{CT}
To ensure renormalizability and preserve tree-level minima, when it deviates even after adding zero temperature 1-loop Coleman-Weinberg potential, we have to add the so-called counter term potential ($V_\text{c.t.}$). Recall $V_0$ from Eq.(\ref{V0}) and  write $V_\text{c.t.}$ in our case as \cite{Taramati:2025ygs}
\begin{equation}
\label{Vct}
    V_\text{c.t.}(h)= -\frac{\delta\mu_h^2}{2} h^2 + \frac{\delta\lambda_h}{4} h^4.
\end{equation}
  All the coefficients of the above equation can be determined by imposing minimization conditions on  the zero-temperature effective potential with counter terms
  \begin{eqnarray}
      \partial_h \left( V_\text{c.t.} +V_\text{CW} \right) |_{h=v} &=0,\\
    \partial_h^2 \left( V_\text{c.t.} +V_\text{CW} \right) |_{h=v} &=0.
  \end{eqnarray}
  After solving these two equations, we can express the two coefficients as
  \begin{eqnarray}
      \delta \mu^2_h =\frac{1}{2}\left( \frac{3}{v}\partial_h V_\text{CW}-\partial^2_h V_\text{CW} \right)\Bigg|_{h=v},\\
      \delta \lambda_h =\frac{1}{2 v^2}\left( \frac{1}{v}\partial_h V_\text{CW}-\partial^2_h V_\text{CW} \right)\Bigg|_{h=v}.
  \end{eqnarray}
  
\subsection*{Daisy coefficients}
\label{subsec:D}
 The leading contribution at  high temperature ($T \gg M_i$) comes from the finite temperature 1-loop potential, i.e., from Eq. (\ref{vth}) up to $\mathcal{O} (T^2)$, is expressed as \begin{equation}
     V_\text{th}= \frac{T^2}{24}\sum_i n_i M_i^2.
 \end{equation} The daisy coefficients for the relevant scalars are calculated as \begin{equation}
     d_{ij}= \frac{1}{T^2} \frac{\partial^2 V_\text{th}}{\partial x_i \partial x_j}.
 \end{equation} 
So, the Debye or thermal mass corrections for all scalars of our model are given by
\begin{align}
\Pi^2_h(T)&=T^2 \left( \frac{\lambda_{h1} + \lambda_{h2}+\lambda_{h3}}{12} +\frac{\lambda_{h}}{2}+\frac{3g^2}{16} +\frac{g^{\prime 2}}{16} +\frac{y_t^2}{4}+ \frac{1}{12}\left(\frac{\Delta M \sin^22\theta}{v}\right)^2 \right),\\
\Pi^2_{G_0}(T)&= \Pi^2_{G_\pm}(T)=\Pi^2_h(T),\\
\Pi^2_{\phi_{1}}(T)&=T^2 \left( \frac{\lambda_1 +\lambda_{h1}}{3} +\frac{\lambda_{12}+\lambda_{13}}{6}\right),\\
\Pi^2_{\phi_{2}}(T)&=T^2 \left( \frac{\lambda_2 +\lambda_{h2}}{3} +\frac{\lambda_{12}+\lambda_{23}}{6}\right),\\
\Pi^2_{\phi_{3}}(T)&=T^2 \left( \frac{\lambda_3 +\lambda_{h3}}{3} +\frac{\lambda_{23}+\lambda_{13}}{6}\right),
\end{align}
Note that the masses belonging to the same multiplet have the same thermal corrections \cite{Comelli_1997,Jangid:2025ded}.
 According to Ref. \cite{PhysRevD.45.2933}, only the longitudinal mode of the gauge bosons gets a thermally corrected mass in the high temperature limit. So, the thermal corrections to the electroweak gauge bosons are \cite{Oikonomou:2024jms}
 \begin{align}
     \Pi^2_{W_L}(T)=&\frac{11}{6}g^2 T^2,\\
     \Pi^2_{Z_L}(T)=& \frac{1}{2}(g^2+g^{\prime 2})\left(\frac{h^2}{4}-\frac{11~T^2}{6}\right)+\frac{1}{2} \Delta,\\
     \Pi^2_{\gamma_L}(T)=& \frac{1}{2}(g^2+g^{\prime 2})\left(\frac{h^2}{4}+\frac{11~T^2}{6}\right)-\frac{1}{2} \Delta,
 \end{align}
 where $\Delta^2=(g^2-g^{\prime 2})^2 \left(\frac{h^2}{4}+\frac{11~T^2}{6}\right)^2+(\frac{gg^\prime}{2}h)^2 $.

%%%%%%%%%%%%%%%%%%%%%%%%%%%%%%%%%%%%%%%%%%
%%%%%%%%%%%%%%%%%%%%%%%%%%%%%%%%%%%%%%%%%%

\section{Decay width of $\psi^\pm$}
\begin{figure}[H]
    \centering
    \includegraphics[scale=0.4]{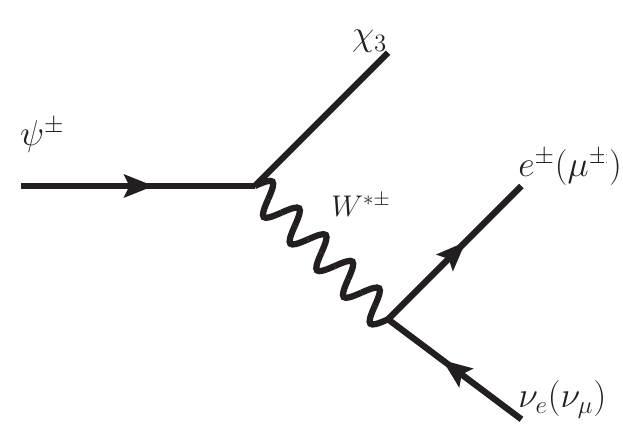}
    \caption{Feynman diagram for three body decay of $\psi^\pm$ to leptons.}
    \label{fig:psi3body}
\end{figure}
The three body decay width of $\psi^\pm$ to DM and charged lepton via off-shell $W^\pm$ exchange, as shown in Fig. \ref{fig:psi3body}, is given as \cite{Cirelli:2005uq}
\begin{eqnarray} \Gamma_{\psi^\pm\rightarrow\chi_3l^\pm\nu_l}=\sin^2\theta\frac{2G_F^2}{15\pi^3}\Delta{M}^5.
\end{eqnarray}

%%%%%%%%%%%%%%%%%%%%%%%%%%%%%%%%%%%%%%%%%%

\providecommand{\href}[2]{#2}\begingroup\raggedright\endgroup

%%%%%%%%%%%%%%%%%%%%%%%%%%%%%%%%%%%%%%%%%%

\end{document}